\documentclass[12pt,preprint]{aastex}

\usepackage[utf8]{inputenc}
\usepackage{graphicx}
\usepackage{captcont}
\usepackage[english]{babel}
\usepackage{ amssymb }

\usepackage{color}

\usepackage[normalem]{ulem}

\slugcomment{}

\shorttitle{Formation of Mars and other terrestrial planets}
\shortauthors{Izidoro et al.}

\begin{document}

\title{Terrestrial Planet Formation in a protoplanetary disk with a local mass depletion: 
A successful scenario for the formation of Mars}

\author{A. Izidoro\altaffilmark{1,2,3} }
\email{izidoro@feg.unesp.br}
\author{ N. Haghighipour\altaffilmark{4,5} }
\email{nader@ifa.hawaii.edu}

\author{O. C. Winter\altaffilmark{1}}

\and

\author{M. Tsuchida\altaffilmark{6}}

\altaffiltext{1}{
UNESP, Univ. Estadual Paulista - Grupo de Dinâmica Orbital \& Planetologia, 
Guaratinguetá, CEP 12.516-410, São Paulo, Brazil}
\altaffiltext{2}{Capes Foundation, Ministry of Education of Brazil, Brasília/DF 70040-020, Brazil.}
\altaffiltext{3}{University of Nice-Sophia Antipolis, CNRS, Observatoire de la Côte d’Azur, Laboratoire Lagrange, BP 4229,
06304 Nice Cedex 4, France.}
\altaffiltext{4}{Institute for Astronomy and NASA Astrobiology Institute, 
University of Hawaii-Manoa, Honolulu, HI 96822, USA }
\altaffiltext{5}{Institute for Astronomy and Astrophysics, University of T\"ubingen, 
72076 T\"ubingen, Germany}
\altaffiltext{6}{UNESP, Univ. Estadual Paulista , DCCE-IBILCE, São José do
 Rio Preto, CEP 15.054-000, São Paulo, Brazil}

\begin{abstract}

Models of terrestrial planet formation for our solar system have been successful in producing 
planets with masses and orbits similar to those of Venus and Earth. However, these models have generally failed to produce 
Mars-sized objects around 1.5 AU. The body that is usually formed around Mars' semimajor 
axis is, in general, much more massive than Mars. Only when Jupiter and Saturn are assumed to 
have initially very eccentric orbits (e $\sim$ 0.1), which seems fairly unlikely 
for the solar system, or alternately, if the protoplanetary disk is truncated at 1.0 AU, 
simulations have been able to produce Mars-like bodies in the correct location. In this paper, we examine an 
alternative scenario for the formation of Mars in which a local depletion in the density of the 
protosolar nebula results in a non-uniform formation of planetary embryos and ultimately 
the formation of Mars-sized planets around 1.5 AU.  We have carried out extensive 
numerical simulations of the formation of terrestrial planets in such a disk for different 
scales of the local density depletion, and for different orbital configurations of the giant planets. 
Our simulations point to the possibility of the formation of Mars-sized bodies around 1.5 AU,
specifically when the scale of the disk local mass-depletion is moderately high (50-75\%) 
and Jupiter and Saturn are initially in their current orbits. 
In these systems, Mars-analogs are formed from the protoplanetary materials that originate in
the regions of disk interior or exterior to the local mass-depletion. 
Results also indicate that Earth-sized planets can form around 1 AU with a substantial amount of water 
accreted via primitive water-rich planetesimals and planetary embryos. We present the results 
of our study and discuss their implications for the formation of terrestrial planets in our solar system.

\end{abstract}


\section{Introduction}

A major obstacle in developing a comprehensive model for the formation of the terrestrial
planets of our solar system is the planet Mars. Despite more than two decades of efforts in
explaining the formation of the inner solar system,(e.g., Chambers \& Wetherill 1998; Agnor et al. 1999; 
Chambers 2001; Chambers \& Wetherill 2001), and many sophisticated and high resolution
computational simulations of the late stage of terrestrial planet formation
(Raymond et al., 2004, 2006; O'Brien et al. 2006; Raymond et al. 2007, 2009),
the formation of Mars is still a mystery. While modern simulations
have been successful in producing a wide variety of terrestrial planets such as two to four planets 
in well-separated and stable orbits (Raymond et al., 2004, 2006; O'Brien et al. 2006; Raymond et al. 
2007, 2009; Lykawka \& Ito, 2013) and have been able to account for the origin of Earth's water and its accretion within the 
timescale consistent with radioactive chronometers  (All\'egre et al. 1995; Yin et al. 2002; 
Touboul et al. 2007; Marty 2012), they have not been able to form a Mars-analog at $\sim$1.5 AU. The body 
that is produced around the current Mars' semimajor axis is, in general, too massive when compared 
to the mass of Mars (See Raymond et. al., 2009 and the review by Morbidelli et al. 2012). 

An important factor in determining the outcome of the simulations of terrestrial planet formation is 
the orbital configuration of giant planets. As shown by many authors, the mean-motion and secular resonances due to these 
objects strongly affect the dynamics of planetesimals and protoplanetary bodies which play a significant role 
in their radial mixing and may also result in their ejection from the system (Chambers 2001; 
Chambers \& Cassen 2002; Levison \& Agnor 2003; Raymond et al. 2004, 2006, 2009; Agnor \& Lin 2012; 
Haghighipour et al. 2013). Regarding the formation of Mars, Mars-analogs with appropriate
masses have only been produced with Jupiter and Saturn 
initially in relatively eccentric orbits (e.g., $e_{\rm Jup}=e_{Sat}=0.1$, Thommes et al. 2008, Raymond et. al. 2009).

Although the current orbits of Jupiter and Saturn do not carry high eccentricities,
early orbital instabilities among these planets (before the beginning of the late stage 
of terrestrial planet formation) could have increased their eccentricities to high values (Lega et al., 2013). 
While such an increase in orbital eccentricity is a natural consequence of planets' instability, the
damping of these eccentricities to their present values, and ensuring that the outer solar system would in fact
develop and maintain its current features, has proven to be a difficult task (Raymond et al. 2009). The high eccentricities 
of Jupiter and Saturn will also affect the process of the delivery of water-rich planetesimals and planetary
embryos to the accretion zone of Earth, causing many of these objects to be scattered out of the system,
leaving Earth with less water-carrying material to accrete. Models of the origin of Earth's water suggest that 
a significant fraction of Earth's water was delivered to its accretion zone
by water-rich planetesimals and planetary embryos from past 2 AU 
(Morbidelli et al., 2000; Raymond et al., 2004; 2006; 2007; 2009; O'Brien et al., 2006; Izidoro et al., 2013). 
An Eccentric Jupiter and Saturn will very quickly deplete the protoplanetary disk inside Jupiter's orbit from its 
water-carrying objects and causes planets formed around Earth's semimajor axis to have much less water than 
expected (Raymond et al., 2009).

To overcome the difficulties of forming a  Mars-sized planet at 1.5 AU, Hansen (2009) proposed that 
terrestrial planets might have formed in a narrow annulus of protoplanetary bodies around 1 AU. The results of 
the simulations by this author showed that they can successfully form Venus, Earth and Mars-analogs in a timescale 
consistent with isotopic constrains [see Kleine et al. (2009) for the constraints on the time of the
formation of Earth, and Dauphas \& Pourmand (2011) for the constraints on the time
of the formation of Mars]. The small mass of Mars in this model is the natural consequence of 
its accretion from the material in the edge of the protoplanetary disk, where it is subsequently
scattered out of the annulus and isolated from other planetary embryos, keeping a low-mass while other planetary embryos grow. 
It is important to note that the formation of Mercury-analog planets is 
still an issue in the models of terrestrial planet formation and was only studied
in a very limited sample of the simulations presented in Hansen (2009).

The success of
the model by Hansen (2009) points to an interesting pathway for the formation of terrestrial planets.
This model requires as much as $2{M_\oplus}$ of material to be confined in the region between 0.7 AU and 1.0 AU
(equivalent to a disk with 3 times the minimum mass solar nebula) implying that a model for the formation 
of the planets in the inner solar system has to be able to present a scenario through which such a large amount
of mass is confined in a small region. 
Recently Walsh et al (2011) proposed that it would be possible to create a narrow annulus of mass around 1 AU 
and also deliver water-carrying objects to the accretion zone of Earth,  a topic that is not considered by
Hansen (2009), if one allows for Jupiter and Saturn to 
migrate inward from their regions of formation to $\sim 1.5$ AU at which point these planets change direction 
and migrate outward to a region close to their current orbits. 
In this model, the delivery of water-rich planetesimals and
planetary embryos to the terrestrial zone is a natural consequence of the 
outward migration of giant planets. Known as the ``Grand Tack'' model, this scenario has been able to
provide the region of terrestrial planets with the necessary amount of mass prior to the onset of their formation. 

Although the Grand Tack model has been able to produce Mars-analogs and account for the current architecture 
of the asteroid belt, it may not agree entirely with the models of the formation and migration of giant planets. 
Hydrodynamical simulations show that the migration  in a protoplanetary disk is very sensitive to numerous 
disk parameters (Morbidelli \& Crida, 2007) which are hard to constrain. In order to assess the validity of 
the Grand-Tack scenario on the formation and migration of Jupiter and Saturn, Pierens \& Raymond (2011)  
simulated the growth of these planets and their dynamical evolution driven by tidal 
interactions with a gas-dominated disk. 
The results of simulations by these authors showed that the two-phase migration of giant planets could be a natural 
outcome of the solar system evolution, probably coinciding with the late phases of the dissipation of the solar 
nebula. In a recent study of the migration of giant planets in an evolved gaseous disk, however, D'angelo \& Marzari (2012) 
have shown that only under favorable conditions and for a small
region of the parameter space, the outward migration of Jupiter 
and Saturn can reach beyond 5 AU (as proposed by the Grand-Tack Model). It is important to note that if the 
outward migration of Jupiter and Saturn is not sufficiently fast, the efficiency of the Grand-Tack model decreases. 
The fast inward-then-outward migration of Jupiter and Saturn is the key to reproducing the architecture of the
asteroid belt, with two distinct populations corresponding to the C- and S-type asteroids. Also, if the
two-phase migration and its change of direction is not fast enough, the protoplanetary bodies in terrestrial 
planet region will be subject to the gravitational perturbation of giant planets for a long time which may result
in their ejection from that region. The latter may subsequently reduce the mass in the planet-forming annulus 
to below its critical value disrupting the formation of terrestrial planets in that region.

 A planet-forming disk is a complex and dynamic environment. During the course of planet formation, beginning from the stage when the dust particles coagulate to 
when the cores of giant planets are formed and terrestrial planets began their accretion, 
the physical properties of the disk are continuously subject to change. For instance, regions may appear where
the disk is turbulent, or at places the density of the gas or solid material may be temporarily enhanced
(e.g., Chiang \& Goldreich 1997; Papaloizou \& Nelson 2003; Laughlin et al. 2004; Garaud \& Lin 2007). 
These processes affect the radial profile and the physical properties of the disk (e.g., temperature
and surface density), and may deviate them from simple power-laws. As proposed by Jin et al. (2008), 
the radial variations in the ionization fraction of the disk, for instance, can separate the disk into 
an inner high-viscosity part and an outer low-viscosity region.
The difference in viscosities in these two regions causes the material in the inner part
of the disk to flow faster toward the star compared to the material in the outer part, creating
a local minimum in the disk's mass distribution in the boundary between these two regions. As 
shown by these authors, during the evolution of the nebula around the Sun, two of such regions might have 
appeared. One region is narrow with a width of approximately 0.1 AU,
varying between 1.3 AU and 3.4 AU, with a most probable location in the range of 1.3-2.4 AU. 
The second region is larger, with a width of $\sim 1$ AU centered at about 1.6 AU.  Jin et al. (2008) 
suggested that the appearance of such a local minimum in the disk 
could result in a non-uniform formation and distribution of planetary embryos around the orbit of Mars, and will 
ultimately result in the formation of Mars-sized objects.

As previously discussed, Hansen (2009) and Walsh et al. (2011) showed that the key 
to the successful formation of Mars is to accumulate planet-forming material
in a narrow region with an outer  edge at 1 AU. 
As a consequence of limiting the protoplanetary disk to such a small annulus, the accretion and
growth of an object will stop when it is scattered outside this region.
The latter seems to present a pathway to the successful
formation of Mars (or, Mars-analogs). In that respect, considering a local 
depletion in the protoplanetary disk around the Mars' location presents an alternative 
to the Grand Tack model for forming a Mars-analog using the Hansen-style 
disk truncation (Hansen, 2009) without the brief migration of Jupiter and Saturn in a gas-rich phase 
as used by Walsh et al., (2011). 
Another important aspect of this scenario is that unlike the model by Hansen (2009) which does not
track the delivery of water to the terrestrial zone due to its initial adhoc set up of the protoplanetary
disk, it presents a water-delivery mechanism that is also different from that of Walsh et al. (2011)
who suggested that the water is delivered to around 1 AU very early during the outward migration of 
Jupiter and Saturn.

In this paper, we examine this scenario by simulating the formation 
of terrestrial planets in a disk with a local minimum in surface density as proposed by 
Chambers \& Cassen (2002) and Jin et al (2008). Our goal is to determine 
the range of the parameters for which a protoplanetary disk with a locally depleted region 
can form a Mars-sized planet around 
1.5 AU, and also produce a planetary system (fairly) consistent with the inner solar system
(i.e., with Venus- and Earth-sized planets and with appropriate amount of water on Earth).  
We consider the local minimum to be around Mars' semimajor axis and use a non-uniform distribution of 
planetesimals and planetary embryos to simulate the final stage of 
terrestrial planet formation. Because during the evolution of the disk, the location of the local minimum 
may vary, we will carry out simulations for different values of the location and depth of this region.

It is important to note that according to Jin et al. (2008), the inward flow of the material from outer 
regions of the nebula to its inner parts due to viscosity differences might have affected the composition 
of the disk material prior to the accretion of terrestrial planets. For instance, volatile-rich dust grains 
or planetesimals could have drifted inward much earlier than the onset of terrestrial planet formation, 
and have played an important role in the final mass and composition of these objects. The study of the 
changes in the properties of the nebula due to the inward flow of material and appearance of the gap is a 
complicated task that has not been accounted for in any rigorous ways in current protoplanetary gas-disk 
models. Such a modeling is also beyond the scope of this paper. In this study, we avoid these complications 
by focusing solely on the last stage of the accretion of planetesimals and planetary embryos. For simplicity, 
we also neglect the re-accumulation of the material that was originally removed from the depleted region of 
the disk during the evolution of the nebula, in particular the material that might have drifted inward into 
the terrestrial planet region. Due to the lack of a proper model for the evolution of the nebula, an initial 
condition for planetesimals and planetary embryos that takes into account such a re-accretion of the drifted 
material would be poorly constrained. 

We describe our model in section 2 and present the results of our simulations in section 3. 
Section 4 concludes this study by summarizing the results and discussing their implications for
the formation of terrestrial planet in our solar system.

\section{The Model}

As shown by Kokubo \& Ida (1998, 2000), during the formation of terrestrial planets, the Runaway and Oligarchic
growths of planetesimals and planetary embryos result in a bi-modal distribution of mass in the protoplanetary disk. Since in this study,
our focus is on the late stage of terrestrial planet formation, in order to be consistent with the bi-modality of the disk,  
we consider a protoplanetary disk of planetesimals and planetary embryos with half of its mass from the planetesimals and the other
half from planetary embryos. The disk extends from 0.5 AU to 4 AU. In such a bimodal disk, the planetesimals provide dynamical friction
which is necessary to damp the eccentricities and inclination of the planetary embryos embedded in the disk (O'Brien et al. 2006; 
Morishima et al., 2008).
The individual planetesimals are considered to have a mass of 0.0025 Earth-masses and distributed with a surface 
density profile of $r^{-3/2}$. These objects only interact with the planetary embryos and the giant planets, and 
do not see each other. 
The masses of the planetary embryos scale as $M\sim r^{3/2(2-\alpha)}\Delta^{3/2}$ (Kokubo \& Ida 2000; Raymond et al. 2005, 2009) 
where $\alpha$ is a free parameter and $\Delta$ is the number of mutual Hill radii. 
We consider the embryo-to-planetesimal mass-ratio to be $\sim 8$ 
around 1.5 AU (Raymond et al. 2009), and the disk surface density to have a radial profile of $r^{-3/2}$ 
(i.e., $\alpha=3/2$) and be given by

\begin{equation}
\Sigma(r)=
\left\{
\begin{array}{lll}
\Sigma_1 ({r}/{1 {\rm AU}})^{-3/2} \qquad\qquad\quad; \hspace{0.4cm} {\rm outside\,\, the\,\, depleted\,\, region,}\\  \\ 
(1-\beta)\Sigma_{1} ({r}/{1 {\rm AU}})^{-3/2 } \qquad; \hspace{.3cm} {\rm inside\,\, the\,\, depleted\,\, region.}
\end{array}
\right.
\end{equation}

\vskip 10pt
\noindent
In this equation, ${\Sigma_1}=8 \, {\rm g/cm^2}$ and the parameter $0<\beta\leq1$ represents the scale of local 
mass-depletion. Figure 1 shows the distribution of planetary embryos and planetesimals for $\beta=50\%$. We chose the location
of the local mass-depletion according to the model presented by Jin et al. (2008). However, because as shown by these
authors, this location changes in time, in order to better explore the parameter space of the system, we considered two different regions of mass-depletion; one extending from 1.1 AU to 2.1 AU denoted as the disk model A, and one extending
from 1.3 AU to 2.0 AU denoted as the disk model B. Table 1 shows these regions and their corresponding parameter $\beta$. 
 We assume that in the beginning of our simulations, Jupiter and Saturn are fully formed, and carry out 
simulations considering different orbital configuration of these planets.

We would like to recall that our goal is to present a model capable of forming terrestrial planets, in particular Mars,
consistent with the inner planets of our solar system. Our approach is to follow the idea presented by Hansen (2009)  
(in which terrestrial planets are formed in a truncated disk of protoplanetary bodies), however, instead of appealing to the 
giant planet migration to create a truncated disk as in the Grand Tack scenario (Walsh et al. 2011), we consider the disk truncation
to be a consequence of the disk's natural evolution. In that respect, the protoplanetary disk in the Hansen (2009) model
can be considered as an extreme case of our disk model, extending from 0.7 AU to 1 AU with a depletion factor of $\beta=100\%$ 
for the region beyond 1 AU.
The fact that in our disk model, the region between 2 and 4 AU is initially populated by planetesimals and planetary embryos (Table 1) is 
a distinct characteristic of our model that makes the results more realistic and, unlike Hansen's simulations, allows for
the formation of the asteroid belt and the delivery of water to Earth from water-rich planetesimals and 
protoplanetary embryos (Morbidelli et al., 2000; Raymond et al., 2004).

\section{Numerical Simulations}

We performed a total of 84 simulations considering two different sets of initial orbital configurations for the giant planets. 
As  mentioned before, we assumed that at the beginning of the simulations these planets were fully formed  
and that the gas disk had been fully dissipated. In the first set of simulations, 
we considered Jupiter and Saturn to be in their current 
orbits. In the second set, we assumed their orbital elements to be similar to those proposed in the Nice Model, that is,
$a_{\rm Jup}=5.45$ AU, $a_{\rm Sat}=8.18$ AU, and both planets to be initially in circular and co-planar orbits 
(Tsiganis et al., 2005; Gomes et al., 2005; Morbidelli et al., 2005).
Each simulation started with 700-950 planetesimals and 110-190 planetary embryos, all initially in circular orbits.
The planetary embryos were spaced from one another at distances of 3-6 mutual Hill radii (Kokubo \& Ida, 2000). 
The initial orbital inclinations 
of all bodies were chosen randomly from the range of $10^{-4}$ to $10^{-3}$ degrees, and their mean 
anomalies were taken to be between $0^\circ$ and $360^\circ$. The arguments of periastrons and longitudes of ascending 
nodes of all objects were initially set to zero. For each value of the depletion parameter $\beta$ and giant planet
configuration, we considered at least three different, randomly generated initial conditions for 
planetesimals and planetary embryos. In view of the stochastic nature of this kind of 
simulations, we performed additional simulations using those parameters that produced planetary systems with features 
close to those of the solar system. We also used the water distribution model by Raymond et al. (2004) and considered 
planetary embryos and planetesimals inside 2.0 AU to be initially dry, the ones between 2 AU and 2.5 AU to carry 0.1\% water, 
and the water contents of objects between 2.5 AU and 4 AU to be 5\%.

Using the hybrid integrator  of the N-body integration package MERCURY (Chambers 1999), we integrated
our systems for 500 Myr. The time-steps of integrations were set to 6 days. At the end of each simulation, we
identified those that produced a potential Mars- and/or an asteroid belt analog, and integrated them for another 500 Myr.

We would like to emphasize that this time of integration is considerably larger than the time 
of the simulations of terrestrial planet formation that is usually found in the literature. 
In general, most of the simulations of the late stage of
terrestrial planet formation are carried out for 100 to 200 Myr (e.g., Chambers 2001, Raymond et al. 2004, 
2006,2009, O'Brien et al. 2006, Hansen 2009, Walsh et al. 2011). We decided to carry out simulations for longer times because
we noticed that after 200 Myr of integration, several of our simulations produced small bodies (including two or three 
potential Mars-analogs) in orbits interior to 2 AU. To examine whether these systems
would be stable, we extended the integrations to longer times.

 Except for planetesimal-planetesimal interactions, we allowed all objects to interact with one another and collide. 
It has been shown by Kokubo \& Genda (2010) that the assumption of inelastic 
collisions in simulations of terrestrial planet formation  has no important effect on the mass and orbital assembly
of the final planets (if a collision results in fragmentation, the remaining material will be accreted in subsequent 
collisions). In a recent study, Chambers (2013) also found that in the simulations of terrestrial planet 
formation where fragmentation is considered, the final planetary systems are broadly similar to those of the simulations 
in which fragmentation is neglected. We therefore considered collisions to be perfectly inelastic, resulting in 
the complete merging of the two colliding bodies while conserving linear momentum.

\section{Results}

In analyzing our results, we define an Earth-analog as a planet that is formed in the region of 0.75-1.25 AU 
and has a mass equal to or near that of Earth. Similarly, a Mars-analog is defined as a 
planet with a mass equal to or near that of Mars ($\sim 0.3 M_{\oplus}$), formed between 1.25 AU and 2.0 AU. 
We also define a terrestrial planet as an object larger
than 0.025-Earth masses ($\sim$0.5 Mercury-mass) with a semimajor axis smaller than 2 AU. 

As mentioned in section 3, we carried out simulations for two different orbital architectures of Jupiter and Saturn
(their current orbits as well as when they are initially in circular orbits as in the Nice model), and for different values of the
scale and location of the local mass depletion. In the following, we present the results of each of these simulations,
and compare the results with the current state of the inner planets of the solar system.

\subsection{Jupiter and Saturn in their current orbits}

The first analysis of our results in this case indicates that in close to $50\%$ of our simulations, 
the final planetary system
includes at least one object with a mass smaller than half of the mass of Earth around 1.5 AU. Tables 2 and 3 show the statistics for this 
body in its final planetary system. Results in Table 2 correspond to the simulations of the disk model A (Table 1) and those in
Table 3 correspond to the disk model B. The roman numbers in the first column of these tables refer to the first, second, and third
set of initial conditions in the simulations for a given mass-depletion coefficient.

Figure 2 shows the snapshots of the dynamical evolution of a sample system corresponding to a simulation
using the disk model A (Table 1) and  for a depletion scale of 75\% (simulation A-75\%-II in Table 2). 
As shown by the bottom-right panel, integrations resulted in the formation of
four planets in the region between 0.5 AU and 1.5 AU with the masses of (from left to right) 0.21, 0.95, 0.55 and 0.08 $M_\oplus$, 
respectively. The color of each body represents its water-mass fraction. The least massive planet at 1.5 AU
is an interesting low-mass object that has reached 90\% of its mass in less than 2.5 Myr. The relatively fast 
formation of this body is in agreement with the timescale of the formation of Mars as suggested by Nimmo \& Kleine (2007) and 
Dauphas \& Pourmand (2011). As argued in the latter article, isotopic analyses require that Mars to have been accreted 
rapidly and reached approximately half of its present size in only $1.8^{+0.9}_{-1.0}$ Myr.
 
Figure 3 shows the results of a simulation for similar mass-depletion scale and initial orbital configuration of
Jupiter and Saturn as in Figure 2, but using the disk model B (simulation B-75\%-III in Table 3). As shown by the bottom-right panel, 
in this simulation, three terrestrial planets are formed with masses equal to 0.7, 0.98, and 0.06 $M_\oplus$. Given the masses and
distances of these planets to the central star, this system shows a close similarity to the Venus-Earth-Mars configuration 
in our solar system. However, unlike the results of the simulation of Figure 2, the planet around 1.5 AU in Figure 3 accreted 
more slowly and took longer than 80 Myr to reach  90\% of its current mass. 
 
Figure 3 also demonstrates a case for which the extension of integrations to 1 Gyr was necessary. In this simulation, 
as shown by the panel corresponding to 200 Myr, a moderate number of small bodies, with two potential Mars-analogs, were formed with orbits between 
1.0 AU and 2.0 AU. The extension of the integration 
to over 500 Myr revealed that only one of these bodies was stable, and the other one was ejected from the system after a 
close-encounter with the proto-Earth at $\sim 530$ Myr. This is a typical outcome that was obtained 
in several of our simulations. In systems with high scales of depletion, in particular,
the depleted region of the disk is initially populated by a large number of small bodies. If these objects are sufficiently 
small, their mutual gravitational interactions will be weak, and as long as no external perturbation affects 
their dynamics, their orbits will stay stable for long times. However, if the sizes of these objects allow for their actual
interactions to be strong, they may scatter each other, in which case the orbits of some of these bodies may become unstable.

One characteristic of the two Mars-analogs shown in Figures 2 and 3 is their moderate orbital inclinations 
relative to the orbit of the largest body in their systems. We note that we
compute orbital inclinations relative to the orbit of the largest planet in the inner part of the system in order to be able to
compare our results with those in Hansen (2009). Unlike the orbit of Mars that 
has an inclination of $\sim 2^\circ$ with respect to the orbital plane of Earth, the two Mars-analog planets in Figures 2 and 3 
have inclinations in the range of 
2$^\circ$-13$^\circ$ with respect to the plane of the orbit of the largest planet in the region between 
0.5 AU and 1.5 AU. Our simulations show that moderate inclinations appear mainly for
planets smaller than 0.1 $M_\oplus$, probably due to the weak action of the dynamical friction on these bodies. 
This enhanced inclination 
has also been reported by Hansen (2009) in his simulations of the formation of terrestrial planets where this author 
showed that Mars analogs are formed with similar inclinations in almost 50\% of the cases. In our simulations, the moderate 
inclinations of planetesimals and  planetary embryos between 1.5 AU and 2.0 AU can be attributed to the effect of the
$\nu_{16}$ resonance. This resonance, as well as the secular resonance $\nu_6$ occur when the periods of the nodal 
and apsidal precessions of the orbit of a small body become equal to those of Saturn. As shown 
in Figure 4, the $\nu_{16}$ resonance significantly increases the inclinations of objects within the first 10 Myr. 

In addition to the $\nu_{16}$ resonance, Figures 2 and 3 also show the strong effects of 
the mean-motion resonances with Jupiter as well as the effect of the $\nu_6$ secular resonance 
with Saturn (see the upper right panel in each figure). As shown here, 
these resonances increase the eccentricities of objects in their vicinities causing many of these bodies to be 
either scattered out of the system, or collide with the Sun, Jupiter, or Saturn.
(Gladman et al., 1997, Levison \& Agnor 2003, Raymond et al. 2006, Haghighipour et al. 2013). 
Although these processes remove a significant 
amount of water-carrying objects from the protoplanetary disk (see middle-right panels of Figures 2 and 3),
simulations indicate that despite a depletion scale of 75\%,
terrestrial planets can still form around 1 AU and carry a significant amount of water. As a point of comparison,
The lower limit of the amount of the Earth's water is 5$\times10^{-4}$ 
of its water-mass fraction (Raymond et al., 2004). The planets around 1 AU in Figures 2 and 3 
carry higher amounts.

\subsubsection{The effect of the scale of mass-depletion}

As expected, the growth of planetary embryos in the mass-depleted region is proportional to the amount of mass 
that is available in that region. In other words, it depends on the depletion factor $\beta$. For large values
of $\beta$ (e.g., 75\% which corresponds to a total initial mass of  $\sim0.25 M_\oplus$ in the region of 1.3-2.0 AU
in disk model B), the small amount of mass in the depleted region combined with the effect of $\nu_{6}$ and $\nu_{16}$ 
resonances has a negative impact on the rapid formation of Mars-analogs during the first few Myr of 
integration. The increase in the orbital inclinations causes planetesimals
and planetary embryos to be in different orbital planes, decreasing the rate of their collisions 
and consequently the efficiency of their growth.
The orbital excitation of these bodies (as a result of both increasing their eccentricities and inclinations),
causes many of these objects to be scattered out of the system devoiding the region from material necessary
to form small planets and Mars-analogs (Haghighipour et al., 2013). For smaller values of $\beta$ where the mass
of the depleted region is larger (e.g., 20\% which corresponds to a total initial mass of  $\sim0.82 M_\oplus$ 
in the region of 1.3-2.0 AU in disk model B), the collision and growth of planetary embryos results in the 
formation of larger objects. For instance, simulations with $\beta = 20\%$ routinely produced planets around 1.5 AU up to
5 times larger than Mars. This result is consistent with those in Raymond et al. (2009).

The results of our simulations also show that embryos that originated in the non-depleted region of the disk 
and were scattered into the depleted region seem to have a better chance of growing to the Mars size and 
maintaining long-term stability. These objects are considerably larger than the planetary embryos 
native to the depleted region and less responsive to the perturbing effects of $\nu_{6}$ and $\nu_{16}$
resonances. The latter is primarily due to the effect of the dynamical friction which strongly damps the 
eccentricities and inclinations of these objects. For instance, both Mars-analogs of Figures 2 and 3
were formed at larger distances ($\sim$ 2.7 AU and 3.1 AU, respectively) and were scattered into the region 
of $\sim$1.5 AU as a result of successive close-encounters with other planetary embryos and the effect of resonances 
(Figure 5). 

The formation of Mars-analogs outside the region around 1.5 AU and their
subsequent scattering into this region has also been observed in the simulations by Hansen (2009) and Walsh 
et al. (2011). As shown by these authors, in those simulations in which a Mars-analog was formed, this object
was accreted in the region around 1 AU and was scattered to the vicinity of the Mars' orbit 
at 1.5 AU. In that region, the object maintained an isolated orbit, and as a result, 
keeping a low mass.

In our simulations, 
Mars-analogs are formed both in the region interior to 1.5 AU (e.g., around 1 AU) where they are
scattered outward, and exterior to this region at 2.0-2.5 AU, where their interactions with other embryos and
resonances causes them to scatter inward. This is a significant result in the sense that it is consistent with
the formation of Mars as a stranded planetary embryo, and implies that depending on the original
region of the accretion of this object, its internal composition will be different. Those Mars-analogs that are
scattered inward  from the regions beyond 2.0-2.5 AU appear in 25\% of our simulations and in general
carry more water than those scattered outward from the inner part of the protoplanetary disk (the remaining 75\%). Depending on the
regions of their origin, these objects will also have different D/H ratios than that of Mars
(Drouart et al. 1999, Morbidelli et al. 2000). At the moment, the primordial value of the D/H 
ratio of Mars' water is unknown (Lunine et al., 2003), however, evidence from Martian meteorites suggests 
that the initial D/H value for Martian interior water is consistent with a chondritic D/H composition 
(Usui et al., 2012). The formation of Mars as an 
inward-scattered planetary embryo can also explain the higher D/H ratio for Mars's water relative to 
that of Earth (Lunine et al., 2004).

The scale of depletion has also an important effect on the radial mixing of 
the bodies and the material-content of the final planets. Radial mixing is driven by the mutual interactions
between protoplanetary bodies as well as the perturbations due to giant planets. Embryo-embryo and embryo-planetesimal
interactions increase the eccentricity of these objects and scatter them into other regions of the disk. Mean-motion
and secular resonances with Jupiter and Saturn (e.g., in the range of 1.8 AU to 3.7 AU when these planets
are in their current orbits) also play important roles, in particular in the delivery of water-carrying asteroid
to terrestrial planets (Raymond et al., 2004). Our simulations indicate that when the depletion scale is low, 
the embryos inside the depleted region are initially larger and as a result, have stronger interactions with their 
neighboring embryos and 
planetesimals. In this case, radial mixing is more efficient and the material-contents of the final planets show
greater diversity in their origin. 

Figures 6 and 7 show samples of the results. Figure 6 corresponds to the simulations using the disk model A,
and Figure 7 shows the results of the simulations using the disk model B. The size of each body 
in these figures corresponds to its relative physical size scaled as $M^{1/3}$. However, it is not to scale on the $x$-axis. 
The color of each object represents the relative contribution of material from different parts of the disk. 
The eccentricity of each planet is represented by a horizontal line corresponding to its variation 
in heliocentric distance over the semimajor axis. As shown in these figures,
in systems with low mass-depletion scales, the water-mass fraction of the final bodies are larger
indicating that in such systems, the final terrestrial planets tend to carry more water (left plots in Figure 8).
As shown in Figures 6-8, when the scale of depletion is 
100\%, terrestrial planets are formed interior to $\sim$1.3 AU with
little or no amount of water. However, a scale of depletion of 75\% or smaller produces Earth-analogs with mean water-mass 
fractions consistent with the value expected for Earth (Table 3).

\subsubsection{The effect of the location of the mass-depletion}

As mentioned earlier, we carried out simulations for different locations of the disk local mass-depletion.
Results indicated that the position of the inner edge of this region plays a significant role in the 
mass and orbital assembly of the final terrestrial planets. A comparison between the results shown in Figures 6
and 7 indicates that for instance, in simulations where the depleted region started at 1.1 AU (Figure 6), the planets 
formed around 1 AU were consistently smaller than Earth. However, in simulations of Figure 7, where the inner edge of 
the depleted region is at 1.3 AU, these planets are relatively larger. This can be explained noting that for similar 
mass-depletion factors, in the simulations of Figure 6 (using the disk model A), the amount of the mass 
available for accretion by embryos around 1.0 AU is smaller and as a result, the final planets have lower masses compared 
to those in Figure 7 where simulations were run for the disk model B.  

Figure 9 shows the mean planet mass for all our
simulations. The top panel in this figure includes all the planets in the system whereas in the bottom panel only those 
formed in the region between 0.75 AU and 1.25 AU have been included. As shown by this figure,  
the disk model B is more efficient in forming Earth-sized planets. It is interesting to note that in contrast to our model, 
Hansen (2009) and Walsh et al. (2011) used a narrower disk that was truncated at 1 AU 
and were still able to form Earth-sized planets around that region. However, in their disk models, 
the initial amount of mass between 0.7 AU and 1 AU is ${\rm 2\, M_\oplus}$ which is almost twice higher than the amount of mass 
between 0.5 AU and 1 AU in our protoplanetary disks.

The outer edge of the depleted region (at 2.0 AU or 2.1 AU, see Table 1), did not seem to have a strong 
influence on the mass and orbital architecture of the final terrestrial planets. However, this boundary 
may play an important role in the efficiency of the delivery of water-carrying objects 
to the Earth's accretion zone. A mass-depleted region extending to distances well beyond 2 AU  (e.g., along all the 
asteroid belt) can hinder the delivery of water to Earth.
We carried out simulations considering a depleted region extending from 1.5 AU to 2.5 AU and assuming Jupiter and Saturn 
initially in their current orbits. Results showed that for depletion scales of 50\% and 75\%, simulations produced 
Earth-analogs that were mainly dry.

\subsection{Jupiter and Saturn in Circular Orbits}

To examine the role that the current orbital eccentricities of Jupiter and Saturn play
in the formation of a Mars-analog and the final assembly of terrestrial planets, we also carried out simulations
assuming Jupiter and Saturn to be initially in circular orbits. We considered the initial orbital elements of these
planets to be identical to those in the Nice model (Tsiganis et al. 2005) and carried out simulations for 1 Gyr. 
Figure 10 shows the results of one of such simulations. As shown here, four planets are formed with semimajor axes smaller 
than 2 AU. Among these planets is an object similar to Mars with a mass of $\sim 0.24 M_\oplus$ at $\sim$1.35 AU.
Although the formation of this planet can be considered as a success in forming Mars-analogs, there is 
another terrestrial planet in this system with a mass of $0.7 M_\oplus$ in a stable orbit at 1.7 AU
which has made its final planetary configuration inconsistent with the current architecture of the solar system.

The lack of consistency between the final planetary orbits and the current orbits of terrestrial planets 
was observed in the results of all simulations in which Jupiter and Saturn were initially in circular orbits. 
Figure 11 shows the mass and semimajor axes of the final bodies. As shown here, in general, 
results are very different from those of the previous simulations where Jupiter and Saturn were initially in their 
current orbits (Figures 2, 3 and 15). Unlike those simulations, when giant planets were assumed to be in circular orbits,  
the final planetary systems did not contain a Mars-sized planet around 1.5 AU. This was   
irrespective of the scale of the mass-depletion as well as the disk model. Results show 
that in all these simulations, only one planet with a mass smaller than $0.3 M_\oplus$ was formed around 
1.5 AU (Figure 11). 

The lack of success in forming Mars-sized planets around 1.5 AU in the simulations in which Jupiter and Saturn
were initially in circular orbits is in contrasts with the results of the simulations by  
Walsh et al. (2011) who also considered Jupiter and Saturn to be in circular orbits and were still able
to produce Mars-analogs. The reason for this discrepancy can be found in the mass-distribution in the disk models
used by these authors. At the end of the inward migration of Jupiter and Saturn,  
these authors set the total initial mass available for accretion in the terrestrial zone 
to $\sim 2 \, M_\oplus$. They also considered that 
the region of the accretion of terrestrial planets is confined to a narrower region from $\sim$0.7 AU to 1 AU. 
After the inward-then-outward migration of Jupiter and Saturn, 
the total mass of the remaining planetesimals and protoplanetary embryos orbiting from 1.5 AU to 4 AU, in their 
model becomes very small. As a result, these objects do not contribute significantly
to the growth of the planetary bodies in the terrestrial zone. In our simulations, however 
there always exist a significant amount of mass ($\sim1.5 \, M_\oplus$) beyond 2.5 AU.  
When the orbits of Jupiter and Saturn are considered to be 
circular, their interactions with the protoplanetary disk are minimal and as a result, the disk maintains 
a large portion of its original mass. Consistent with previous studies in which Jupiter and Saturn were assumed to be in circular orbits 
(e.g., Wetherill 1996; Raymond et al. 2005, Kokubo et al. 2006), a more massive disk in these simulations produces more 
massive objects. Figure 12 shows the mean mass of the planets formed in our simulations considering different giant planet configurations.

The smaller perturbations of the giant planets on the protoplanetary disk in simulations in which these planets 
are considered to be in circular orbits also affects the 
water contents of the terrestrial planets. In these simulations, the radial mixing of the 
protoplanetary bodies is more effective and results in forming planets with higher contents of water 
(Figure 8, see also Raymond et al. 2009). 
That is because more water-carrying objects from the outer part of the protoplanetary disk 
maintain their orbits for longer times. 
A comparison between the result of the simulations shown in Figures 3 and 10, which correspond to the same
disk model, indicates that in the simulations where Jupiter and Saturn are in their current orbits, the protoplanetary disk
loses a large portion of its water-carrying objects as a result of the stronger interactions of these giant planets with
the planetesimals and planetary embryos. We expect comparable results to be obtained using an updated version of the Nice model 
(Levison et al., 2011) as well.

\subsubsection{The effect of the scale and location of the mass-depletion}

When Jupiter and Saturn are considered in circular orbits, our simulations do not
show a clear correlation between the results and the different 
values of the scale and location of the mass-depletion (Figures 11-13). As discussed above, 
when the giant planets are in circular orbits, the protoplanetary disk is  
perturbed only weakly, and it maintains a higher fraction of its original 
mass for longer times. The effect of the scale of the 
mass depletion in this case vanishes probably because a significant part of the material in the 
neighborhood of the depleted region enters in 
the depleted area during the evolution of the system. 
This material comes mainly from the outer part of the protoplanetary disk as the gravitational 
effects of Jupiter and Saturn are weak, and only remove a small fraction of the disk bodies.

\section{Comparison with solar system and other simulations}

To compare the results of our simulations with the current orbital architecture of terrestrial planets
in our solar system and with the results of simulations by other authors, we calculated the radial
mass concentration statistics (RMC) and angular momentum deficit (AMD) of the final planetary systems
of our simulations. The value of RMC varies with the semimajor axes of planets and is given by
(Chambers 1998, 2001; Raymond et al. 2009)

\begin{equation} 
{\rm RMC} = {\rm Max}\left(\frac{{\sum_{j=1}^N}\> m_j}{{\sum_{j=1}^N}\> m_j\big[\log_{10}\left(a/a_j\right)\big]^2}\right)\,. 
\end{equation}

\noindent
In this equation, $m_j$ and $a_j$ are the mass and semimajor axis of planet $j$, and $N$ is the number of final bodies. 
The AMD of a system represents a measure of the deviation of the actual 
orbital angular momentum of the planets from the total angular momentum of the system, had the planets been in circular 
and co-planar orbits. Following Laskar (1997), we use equation (3) to calculate this value,

\begin{equation}
{\rm AMD}= \frac{{\sum_{j=1}^N}\Big[ m_j \sqrt{a_j} \> \Big( 1 - \cos i_j \> \sqrt{1-{e_j}^2}\> \Big)\Big]} 
{{\sum_{j=1}^N}\> m_j \sqrt{a_j}}.
\end{equation}

\noindent
In this equation, $e_j$ is the orbital eccentricity of planet $j$, and $i_j$ is its inclination with respect to 
an invariant plane.

Figures 6 and 7 show the values of the RMC and AMD of their corresponding planetary systems.
As a point of comparison, the inner planets of the solar system are also shown. 
We recall that in these simulations, Jupiter and Saturn were initially in their 
current orbits. We would also like to note that, similar to the planet V proposed by Chambers (2007), some of our simulations
produced small planets (mainly unaccreted embryos) past 2 AU. 
However, we did not take these planets into account when calculating RMC and AMD values.

Figures 13 and 14 show the mean values of the RMC and AMD
of our simulations for different scales of the mass-depletion, disk models, and giant planet configuration. The circular solid 
points in these figures represent the means of the RMC and AMD calculated from a set of planetary systems produced 
by at least three different simulations with the same mass-depletion scale. The vertical bars 
on each point represent the lower and upper values of the RMC and AMD in the sample over which their means  
were calculated. The values of the RMC and AMD of Mercury, Venus, Earth and Mars 
(hereafter  MVEM) are also shown in these figures. A comparison between these 
values with the values of the RMC and AMD of the sample results shown in Figures 6 and 7, and the 
mean values shown in Figures 13 and 14 indicates that the mean RMC of our
planetary systems are significantly smaller than that of MVEM. The higher value of the RMC of MVEM is mainly due the orbital 
proximity of these planets and the comparable masses of Venus and Earth.
The lower values of the mean RMC in our results indicate that our simulations form planets in more widely spread orbits
than the separation of the orbits of MVEM.

Our results also indicate that, except for a few planetary systems,
the mean AMD of our simulations is, in general, slightly higher than the AMD of MVEM.
This is mainly due to the initial distribution of the total mass of the disk among planetesimals and planetary
embryos which can be adjusted to result in systems with lower AMDs. As shown by O'Brien el al. (2006), 
it is possible to obtain lower AMDs by increasing the initial value of the planetesimal/embryo mass-ratio
while keeping similar distribution of disk mass between the embryo and planetesimal populations 
(e.g., assigning 50\% of the disk mass to planetesimals and 50\% to the planetary embryos). 
The higher values of the planetesimal/embryo mass-ratio will enhance the dynamical friction which will then be more effective 
in reducing the value of AMD. The dynamical friction can also be enhanced if a  residual population of 
very small objects still exist in the inner solar system after the terrestrial planet formation is completed (Schlichting et al., 2012).

To compare our results with previous studies, we first consider the simulations of the formation of terrestrial 
planets by Raymond et al. (2009). A comparison between the values of the RMC of our systems and those from these authors indicates
that only when in their simulations, Jupiter and Saturn are initially in eccentric orbits ($e \sim 0.1$), the values of their RMC
are similar to those in our best models in which a Mars-analog is formed.
The reason is that in their simulations, the eccentric orbits of
Jupiter and Saturn cause these planets to have strong interactions with the objects in the outer part of the disk, 
removing the majority of them from the system in a short time and creating an edge for the disk at $\sim$2 AU 
(few million years, Raymond et al. 2009). As the material is removed from the disk, 
the $\nu_6$ secular resonance shifts interior to 2.0 AU (Gomes, 1997; Haghighipour et al., 2013) and continues 
to remove material from the inner part of the disk. This causes the disk to develop a local mass-depleted region 
similar to the one considered in our disk models. We note that although the radial mass concentrations in all our simulations 
are smaller than that of the terrestrial planets, in general, they are higher than those in most of the simulations by 
Raymond et. al (2009). 

A comparison of our results with those in the simulations by Hansen (2009) and Walsh et al. (2011) shows 
that the values of their RMC are higher than the average value found in our results. 
The reason can be attributed to the extent of the protoplanetary disks in those studies. Both in simulations by Hansen (2009) and 
Walsh et al. (2011), the total mass necessary for the accretion of terrestrial planets was distributed over a small 
region between  0.7 AU to 1 AU. Such a narrow distribution will naturally result in the formation of planetary systems 
with high values of RMC. Recall that in our simulations, similar to the most of the previous studies of terrestrial 
planet formation in the solar system  (e.g., Raymond et al. 2009), the protoplanetary disk extends from 0.5 AU to 4 AU. 
As shown by our results, it is possible to form planetary systems with higher values of RMC in such disks when
Jupiter and Saturn are considered in their current orbits by using a large mass-depletion scale (Figure 13). 

The results of our simulations indicated that Mars-sized planets have a better chance of forming 
around 1.5 AU in systems where Jupiter and Saturn are initially in their current orbits and when the 
mass-depletion factor has a moderately large value (50\% - 75\%). For the purpose of extending 
our analysis, we identified this subset of our simulations, and following Raymond et al. (2009) 
and recalling our definitions of Mars- and Earth-analogs in Section 4, we evaluated their success, 
quantitatively,  in producing

\begin{itemize}

\item a Mars-analog with a mass smaller than $0.3 {M_\oplus}$ in less than 10 Myr (with half-accretion in less than 2.7 Myr),
\item an Earth-analog with a mass larger than $0.7 {M_\oplus}$ and a water-mass fraction larger than $5 \times {10^{-4}}$ in 30-150 Myr,
\item a system of terrestrial planets with AMD $<$ 0.0036 (twice the MVEM value), and
\item no stranded embryos in the asteroid belt with masses bigger than $0.05 {M_\oplus}$.

\end{itemize}

\noindent
The results are presented in Tables 4 and 5. As shown in these tables,
different combinations of the scale and location of the disk local mass-depletion
can result in systems that in many occasions partially, and in some specific cases almost
entirely satisfy the above-mentioned requirements. When the scale of the
local mass-depletion in the disk is moderately large (e.g., $\beta = 50\%\,-\,75\%$), the final systems
are more successful in meeting the above mentioned criteria, especially when the depleted 
region is considered to be between 1.3 AU and 2 AU (Table 5).

\subsection{Formation of asteroid belt analogs}

One important feature of our solar system that imposes strong constraints on the models of terrestrial planet
formation and solar system dynamics is the asteroid belt. 
Any model for the formation of the inner planets has to also be able to account for the existence of small 
bodies between 2.1 AU and 3.2 AU, and their orbital architecture. In our simulations, several systems showed
signs of asteroid belt analogs. For instance, in the simulations of Figure 3, three planetesimals remained stable
in the region of 2 AU to 3.5 AU, for the duration of the integration (1 Gyr). This simulation also produced an 
Earth-analog ($\sim 1 M_\oplus$) around 1 AU in less than 150 Myr, which is consistent with the timescale of 
the formation of the Earth-Moon system (Jacobsen 2005; Touboul et al 2007). The innermost planet in this system 
is a Venus-analog. 

The final number of planetesimals in the region between 2 AU and 3.5 AU, i.e. their long-term ($>$ 100 Myr) stability, 
depends on the interactions between these objects and their neighboring planetary embryos. The latter itself depends on the
interaction of giant planets with the protoplanetary disk. Figure 15 shows 
the final distribution of surviving bodies in those simulations that had full or relative success in 
producing Mars-analogs around 1.5 AU (see also Table 5). As shown here, many embryos with long-term stable orbits
are formed in the asteroid belt. Quantitatively, 
in 14 simulations out of 18 shown in Figure 15, either no embryo was left in the asteroid belt region, or if 
there was any embryo, its mass was smaller than  $0.05 M_\oplus$. These results agree with the findings of Raymond et al. (2009) 
who showed that in systems where giant planets are in slightly excited orbits, some embryos may maintain their orbits in 
the asteroid belt region for 200 Myr of integration. These authors also showed that it is improbable that during the 
formation of terrestrial planets, a Mars-sized embryo could have been stranded in the asteroid belt and maintained its
orbit for 100 Myr (Raymond et al., 2009; Brasser et al., 2011). Such an embryo would have disturbed the orbits of other 
bodies in its vicinity, creating a gap in the protoplanetary disk that is not observed in the present day asteroid 
belt\footnote{Note that this gap is different from those  that are due to mean-motion resonances with giant planets.}.
Our results also agree with this finding. In regard to the latter, we also
analyzed the results of our simulations where Jupiter and Saturn were initially in circular orbits. We found that planetary 
embryos that survived in the asteroid belt were, in general, much larger than Mars, reaching to a mass 
equal to almost half of the mass of Earth (see Figure 11).

Figure 15 also shows that
in the simulations in which the depletion scale is 50\%, the orbital elements of the
population of asteroids in the region of the asteroid belt are closely
similar to those of the real population of the main belt. 
In the simulations where the depletion scale is 75\%, on the other hand, asteroids were produced mainly between 2.8 AU and 3.2 AU, 
with a few bodies also between 2.2 AU and 2.8 AU. We would like to note that because these simulations are very time-consuming, 
we adopted a moderate resolution. Higher resolution simulations may produce results that may show better agreement with the 
structure of the asteroid belt.

An interesting result depicted by the top panels of Figure 15 is the appearance of a dual-mass population among 
the surviving planetesimals in the asteroid belt. One population is at 0.0025 Earth-masses outside the depleted 
region in both panels. The second population is around 0.00125 Earth-masses in the left panel and 0.000625 Earth-masses 
in the right panel. The appearance of such a dual-mass population is due to the initial assumption of the existence of a local
depletion in the disk. The small objects in the asteroid belt are primarily native to the depleted area 
whereas the larger ones originated outside of the depletion. The native objects are also small 
in number (only 4 in each of the 9 simulations shown in Figure 15).
The small mass and number of the native planetesimals is due to the effect of resonances which causes
many of these bodies to be scattered out of the system. As a result, the asteroid belt analogs that are formed in our
simulations do not carry a mass-gradient distribution.

The small number of the final surviving planetesimals in the region of the asteroid belt in our simulations also points to a
strong clearing process during which many objects were scattered out of the system.
In several of our simulations, the initial number of planetesimals in the asteroid belt was over 400.
However, on average no more than 3 planetesimals survived in the end of the simulations 
(e.g. Figure 2 and 3). This value corresponds to a very small fraction of the initial number of planetesimals
in that region which is in agreement with previous 
studies indicating that a vast majority of the mass of the asteroid belt was removed during the evolution of the solar 
system and the formation of terrestrial planets (Petit el al. 2001)

\section{Conclusion and Discussion}

We studied the late stage of the accretion of terrestrial planets in a disk of 
protoplanetary bodies with a locally depleted region. Following Jin et al (2008), we considered that a
depletion in the density of the protosolar nebula will result in a non-uniform formation of planetesimals and planetary embryos, 
and studied the effects of the scale and location of such a local depletion on the mass, water-content, and final orbital 
assembly of terrestrial planets. For simplicity, we neglected the reallocation/redistribution of mass from the depleted 
region into the inner parts of protoplanetary disk due to the early radial in-flow of material (Jin et al., 2008). This 
enabled us to avoid complications that would rise from considering a more complex compositional and mass-size gradient 
for the initial distribution of solids in our protoplanetary disk models. 

As expected, the final mass and orbital assembly of the planets in our simulations were strongly affected
by the initial orbital configurations of Jupiter and Saturn. The results of integrations showed a clear
distinction between the outcome of simulations in which the orbits of the giant planets were considered to be initially
circular and those in which Jupiter and Saturn had slightly eccentric orbits.
Similar to previous studies (O'Brien et al., 2006; Raymond et al., 2009), when giant planets were
in circular orbits, simulations were systematically unsuccessful in forming planetary systems with Mars-analogs at 1.5 AU. 
In this case, even in simulations with high scale of depletion, 
the effect of the local disk-depletion seemed to vanish and the terrestrial planets that
formed around Mars' orbit were more massive than 0.3 $M_\oplus$. These results suggest 
that if Jupiter and Saturn were initially in circular orbits, as in the Nice Model 
(Tsiganis et al. 2005) or a recent model by Levison et al. (2011),
a disk with a local mass-depletion may be able to form a Mars-sized planet around 1.5 AU only if an additional mechanism
removes material from the disk in the region of the asteroid belt. However, such a mechanism has to also ensure that the
delivery of water-carrying planetesimals to the region of Earth accretion will stay efficient.

When Jupiter and Saturn were placed in their current orbits, the interactions of 
these planets with the protoplanetary disk played an important role in forming Mars-analogs. 
Results of our simulations indicated that in this case,  
a significant portion of the disk material is removed from its outer regions creating a favorable condition for Mars-analogs to 
form around 1.5 AU when the depletion factor has a moderately large value of 50\% to 75\%. In these 
simulations, when considering the disk model B, in addition to forming a Mars-analog, our models were able to deliver 
sufficient amount of water to their corresponding Earth-analogs in $\sim 40\%$ of the cases (Table 5). 
The Mars-analogs in these simulations formed as an embryo that was scattered from the non-depleted region
(either the inner or the outer part) into the depleted area. Simulations showed that a depleted region around 1.5 AU
with a high scale of depletion increases the stability of this planet, although caution has to be taken since larger
values of the mass-depletion will have negative impact on the radial mixing of planetesimals and planetary embryos, and
can deprive Earth from having sufficient amount of water. Embryos scattered from the other parts of the disk to the 
depleted region are substantially more massive than the native planetary embryos, and as a result will have a 
higher chance of being stable. These embryos also tend to grow slowly since not much material will be 
available in the depleted area to accommodate their collisional growth.

The results of our simulations also showed that as expected, the final semimajor axes and eccentricities of 
Jupiter and Saturn were slightly different from their initial values. Such changes in the orbital elements of the giant
planets have also been reported by Chambers \& Wetherill (2001), and Raymond et al. (2004), and are the result of the interaction of these 
bodies with planetesimals and planetary embryos, and the subsequent decrease in the mass of the protoplanetary disk due
to the scattering and ejection of these objects from the system.
In this study, at the end of the simulations in which Jupiter and Saturn were initially in their current orbits, the 
semimajor axis and eccentricity of Jupiter decreased by an average of 0.045 AU and 0.035, respectively.
However, no significant changes occurred to the semimajor axis of Saturn. In most case, the final 
semimajor axis of this planet was within a range of $\pm 0.03$ AU from its initial value. 
The eccentricity of Saturn decreased by 0.035.

The above-mentioned damping of the final eccentricities of the giant planets suggests that as the formation of terrestrial
planets approaches its final stage (or after these planets are fully formed), a mechanism has to exist to ensure that
the orbits of Jupiter and Saturn will have eccentricities similar to their current values.
One approach is to consider the initial orbital eccentricities of these planets, at the beginning of each simulation, to be large. 
Raymond et al. (2009) considered,
for instance, an initial value of 0.1 for the eccentricities of both Jupiter and Saturn, and showed that the final
eccentricities of these planets will be close to their current values (0.05). Whether such large orbital eccentricities
are possible is, of course, a matter of debate. As shown by Lega et al. (2013), early dynamical instabilities between
Jupiter and Saturn that may occur at the end of the lifetime of the gas disk could increase the orbital eccentricities
of these bodies. However, Raymond et al. (2009) have presented a list of arguments as to why Jupiter and Saturn could not
have evolved into such a dynamical state. Results of our study suggest that modest values of orbital eccentricities 
for Jupiter and Saturn are required to ensure the efficient formation of Mars-analogs. Given the subsequent damping 
of the eccentricities of these planets, our model also requires a late-stage dynamical instability between Jupiter and
Saturn, such as that proposed by Tsiganis et al., (2005) to raise the orbital eccentricities of these planets to their current values.

At the beginning of each simulation, to create a low-density region, we scaled down the masses of 
planetesimals and planetary embryos inside the depleted area. However, depending on the scale of
 the mass-depletion, the efficiency of the growth of planetary embryos from planetesimals (Kokubo \& Ida, 2000)
 inside the  depleted area may be so  low that during the Runaway and Oligarch phases, either no planetary embryo 
is formed, or they may  not grow to large sizes. This implies that even if we had not scaled down the masses
 of the objects (e.g. this region was initially populated only by planetesimals), the results would not have changed 
and Mars-analogs would have formed in the same fashion as in our simulations. As mentioned before, Mars-sized objects 
originated from the regions outside the mass-depleted  area and because they had more mass to accrete, they were larger 
than the native bodies. Once inside the deplete area, these objects had a better chance of growing to the Mars’ size because 
they dominated this region gravitationally and accreted (or scattered) smaller, native bodies more efficiently. 

The results of the simulations that were successful in forming Mars-analogs around 
1.5 AU also show that unlike simulations with low or no local mass-depletion (e.g., Chambers 2001; Raymond et al. 2009),
a scale of depletion of 50\% to 75\% is crucial for building planetary systems with stronger radial mass concentrations. 
Although our simulations did not produce RMC values as high as that of the current terrestrial planets, 
their results point to potential pathways for improving models of terrestrial planet formation using a local mass-depletion.
The correlation between the scale of mass-depletion and the RMC value also indicates that the strong radial mass 
concentration of terrestrial planets and the low mass of Mars are two characteristics of the solar system that are deeply connected.

Despite the success of our model in forming planets similar to Mars
and ensuring the delivery of sufficient amount of water to Earth, 
none of our simulations was able to reproduce all the features of the inner solar system. The hardest 
constraints to satisfy was the fast formation of Mars. As suggested by Nimmo \& Kleine (2007) and Dauphas \& Pourmand (2011), 
the measurements of the Hf/W ratio in the Martian mantle point to a timescale
of 0-10 million years for the formation of the core of Mars, and a time of slightly less than 2 million years during which 
Mars reaches to 50\% of its current mass. Although the formation of Mars-analogs around the current orbit of Mars was successful 
in many of our simulations (Table 4 and 5), the time during which these objects grew to 50\% of their masses was in general
 (in more than 92\% of the simulations of Table 4, and in more than 95\% in the simulations of Table 5) much longer than 2 million years.

It is important to emphasize that this longer time of Mars accretion should not be considered 
a weakness for our model. The 0-10 million years timescale for the growth of the core of Mars, as suggested by
 Nimmo \& Kleine (2007), is based on the analysis of a limited number of shergottite–nakhlite–chassignite (SNC) 
meteorites that are assumed to have formed the bulk of Mars mantle. Also, the time of the growth of Mars to 50\% of its 
current mass as suggested by
Dauphas \& Pourmand (2011) is based on the assumption that the mass-evolution of planetary embryos during the oligarchic
state can be modeled by a simple analytic solution if a uniform size of planetesimals is adopted (Chambers, 2006).
However, using N-body simulations, Morishima et al. (2013) have shown that such an analytic approximation for the  
mass-evolution of planetary embryos does not agree with the final masses of these objects at the late stage of the oligarchic growth. 
These authors also showed that for a disk with a minimum mass solar nebula, the timescale for the formation of Mars maybe much 
longer than those derived from the Hf-W chronology -- a result that is consistent with our finding as well.

When comparing the time of the formation of Mars obtained from numerical simulations such as those presented
here, with those obtained from cosmochemical studies, another important factor that needs to be taken into consideration
is that unlike analytical approximations of the mass-evolution of embryos, numerical simulations include the effects of 
Jupiter and Saturn that are assumed to have been fully formed in the beginning of simulations. This is
probably 1-3 Myr after the formation of the first solids (Raymond et al. 2009), the calcium-aluminum inclusions 
[CAIs, dated at 4.568 billion years ago (Bouvier et al. 2007)]. 
As shown by many authors [see e.g., Haghighipour \& Scott (2012) and references therein], these planets
have profound effects on the distribution and growth of planetary embryos during their own formation and after they are fully formed.
The perturbations of these planets cause many embryos to be scattered out of the planet-forming region -- a process that
will have important consequences on the final masses of terrestrial planets. The analytical treatment of the growth of embryos,
however, does not take the effects of giant planets into account.

Numerical simulations of the formation of terrestrial planets such as those presented here
and by Morishima et al. (2013), consistently form Mars at timescales larger than that suggested by 
Nimmo \& Kleine (2007) and  Dauphas \& Pourmand (2011). As argued by Morishima et al. (2013) and Kobayashi \& Dauphas (2013), 
this time can be reduced if Mars grew by the accretion of small fragments and pebbles, or the
surface density of the protoplanetary disk around the orbit of Mars is locally enhanced. This latter scenario seems to be
consistent with the assumption of a local mass-depletion in the proto-solar nebula due to the variation in gas viscosity
as suggested by Jin et al (2009)\footnote{We would like to note that the results presented by Jin et al (2009) and their
suggested mechanism for generating a local mass depletion in the disk depend highly
on the choice of some poorly constrained parameters such as the rotational velocity of the cloud's core and the viscosity of 
the gaseous disk. In fact, regions of local mass depletion may appear during the evolution of a disk for a verity of reasons, and may
have long enough lifetimes to affect the formation and growth of planetesimals, and appear as local mass depletions in 
the protoplanetary disk as well.}. Such a depletion may cause the material to accumulate outside the depleted
area, creating regions where the surface density of the disk is locally enhanced. 
These density-enhanced regions may produce large embryos of the size of Mars 
in short timescales during the Runaway and Oligarchic growth phases 
(Kokubo \& Ida, 2000). In our simulations, Mars-sized bodies do in fact form in the regions
interior or exterior to the local mass-depletion where the disk surface density may be locally enhanced (Jin et al., 2008).
These objects are then scattered into the region of mass-depletion where they maintain a stable orbit for a
long time. This implies that if Mars were a planetary embryo that grew in a density-enhanced region, it could
have formed in a short timescale consistent with the finding of Dauphas \& Pourmand (2011). However,
we recall that as mentioned in Section 1, we do not use such local surface density enhancements when 
generating the initial structure of our protoplanetary at the beginning of our simulations.

One important aspect of our simulations is the treatment of the two-body collisions. In all our simulations, we assumed
that collisions were perfectly inelastic and resulted in the complete merging of the two impacting bodies. We also assumed
that no water was lost during an impct (Marty \& Yokochi 2006) or hydrodynamic escapes (Matsui \& Abe 1986),  
and the total amount of water in the final body would be equal to the sum
of the water contents of the colliding objects. In more realistic simulations, the 
loss of volatile materials during an impact has to be taken into account (Genda \& Abe 2005, Canup \& Pierazzo 2006).

In closing we would like
to emphasize that the works of Hansen (2009) and Walsh et al. (2011) provided significant context and 
motivation for the development of this study. Walsh et al. (2011) combined the narrow-disk idea of 
Hansen (2009) with giant planet migration, and developed a model for the formation of Mars and delivery of water
to Earth in which most of the 
water delivered to terrestrial zone is from primitive asteroids that were scattered inward during the 
outward migration of giant planets. We, however, showed that it would be possible to form a low-mass planet around  Mars' 
semimajor axis (and deliver sufficient amount of water to Earth) 
without considering drastic inward migration for Jupiter and Saturn, if the protosolar nebula did have a 
locally depleted region around the Mars' location. The efficiency of the formation of Mars-sized planets 
around 1.5 AU will be high when the giant planets are considered to be farther out than initially assumed in the 
Grand Tack model (e.g., close to their current locations) and in slightly eccentric orbits (e.g., close to their 
current values). In this model, contrary to what
Jin et al. (2008) proposed, that a Mars-analog would form from the embryos native to the depleted region, our simulations 
show that this object is formed outside the depleted area and is scattered into this region as a result of interacting with
giant planets and other planetary embryos.
Also, although our model uses the same narrow-disk idea as in Hansen (2009) and Walsh et al (2011), our scenario for 
water-delivery to Earth is completely different. In our model, similar to previous studies such as those by Raymond et al. (2009),
water comes from the asteroid belt region due to the perturbing effects of giant planets. This will possibly result in 
the accretion of a 
different fraction of primitive asteroids by the forming Earth, as well as a different time for their accretion. 
 More precise measurements are needed to determine the fraction of the material from different parts of the solar nebula that 
contributed to the formation of terrestrial planets and the origin of the Earth's water. Such measurements will have
profound effects on constraining models of terrestrial planet formation in our solar system and can be used to differentiate
between existing scenarios.

\acknowledgements

We would like to thank the referee, Kevin Walsh, for his very constructive comments that greatly improved 
this manuscript. We are also thankful to Alessandro Morbidelli for his carefully reading of our paper
and his helpful comments. AI and OCW would like to thank Rafael Sfair for his assistance with the 
computing cluster that was used to run part of these simulations. AI and OCW would also like to 
acknowledge financial support from the Brazilian National Research Council (CNPq), Coordenacao de Aperfeicoamento 
de Pessoal de Nível Superior (CAPES),  and  FAPESP - São Paulo State Funding Agency, proc. 2011/08171-3. 
AI wishes to thank the Institute for Astronomy at the University of Hawaii for their kind hospitality 
during the course of this project.
NH acknowledges support from the NASA Astrobiology Institute under Cooperative Agreement NNA09DA77A at the 
Institute for Astronomy, University of Hawaii, and the Alexander von Humboldt Foundation. NH is also thankful to the Computational
Physics group at the Institute for Astronomy and Astrophysics, University of T\"ubingen for
their kind hospitality during the course of this project.

\clearpage
\begin{figure}
\centering
\includegraphics[scale=1]{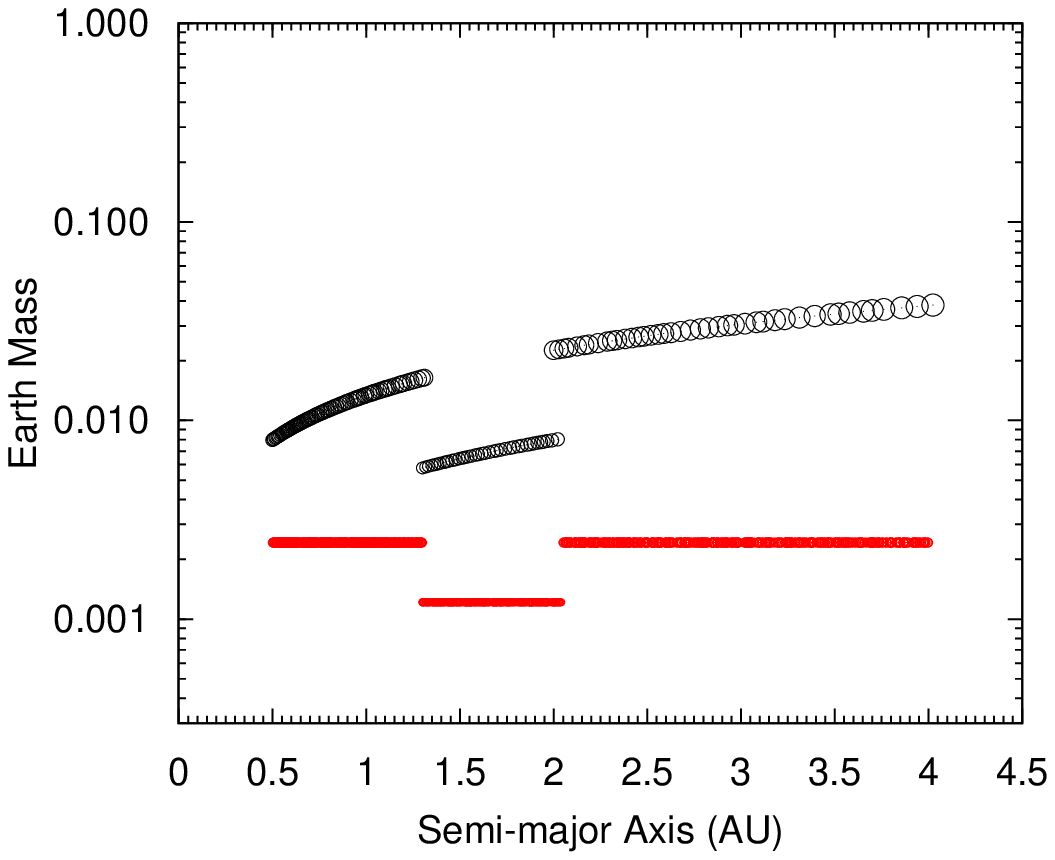}
\caption{Initial distribution of 154 embryos (black) and 973 planetesimals (red) considering a 
mass-depletion of 50\% extending from 1.3 AU to 2.0 AU. The masses of planetesimals are smaller than 
0.003 Earth masses.}
\end{figure}

\clearpage
\begin{figure}
\centering
\includegraphics[scale=0.75]{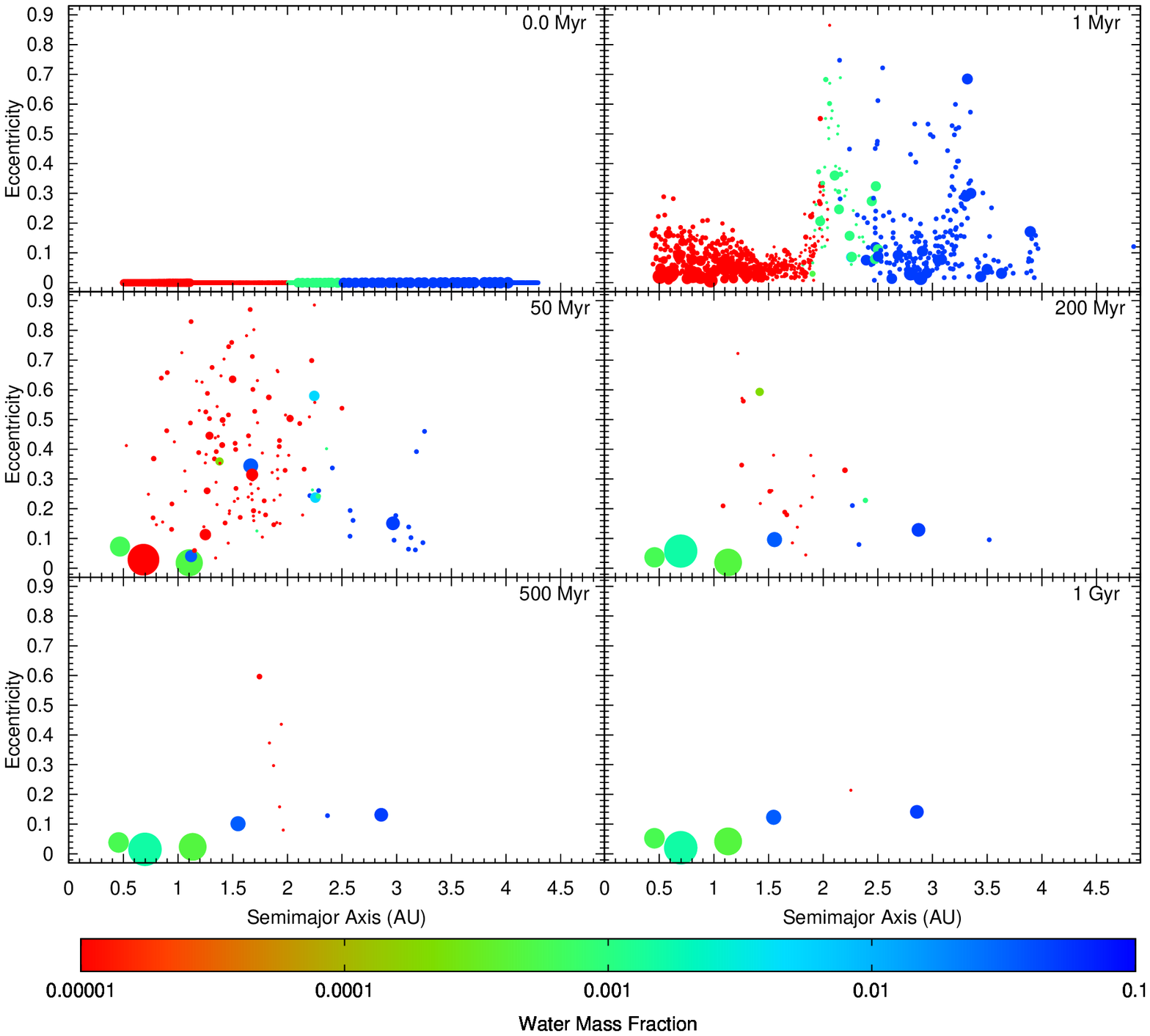}
\caption{Snapshots of the formation and dynamical evolution of planets in a disk with a depletion of 75\%  
extending from 1.1 AU to 2.1 AU. Jupiter and Saturn are in their current orbits. The size of each
body corresponds to its relative physical size and is scaled as $M^{1/3}$. However, it is not to 
scale on the $x$-axis. The color-coding represents the water-mass fraction of the body.}
\end{figure}

\clearpage
\begin{figure}
\centering
\includegraphics[scale=0.75]{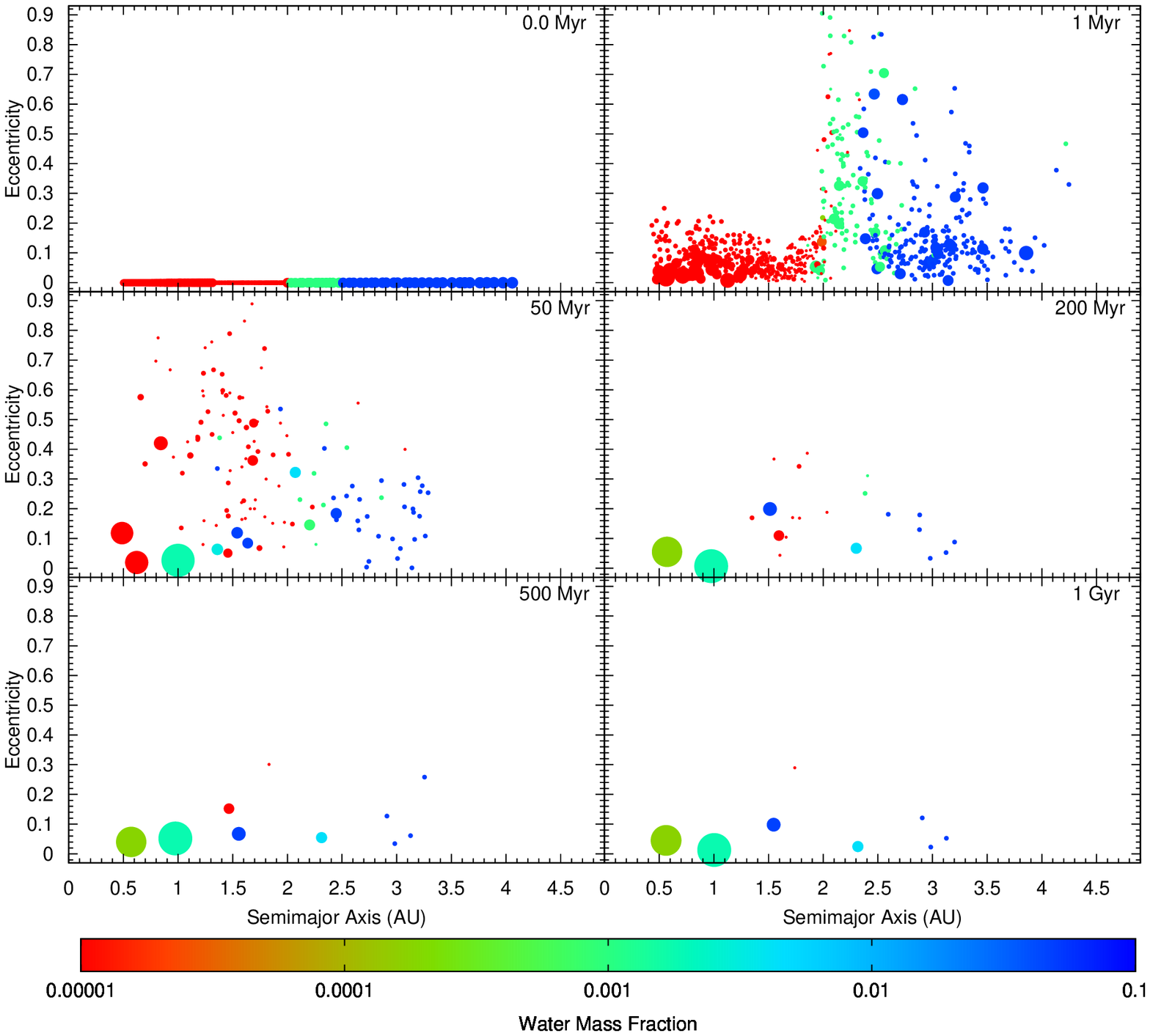}
\caption{Snapshots of the formation and dynamical evolution of planets in a disk with a depletion of 75\%  
extending from 1.3 AU to 2.0 AU. Jupiter and Saturn are in their current orbits. The size of each
body corresponds to its relative physical size and is scaled as $M^{1/3}$. However, it is not to 
scale on the $x$-axis. The color-coding represents the water-mass fraction of the body.}
\end{figure}

\clearpage
\begin{figure}
\centering
\includegraphics[scale=.8]{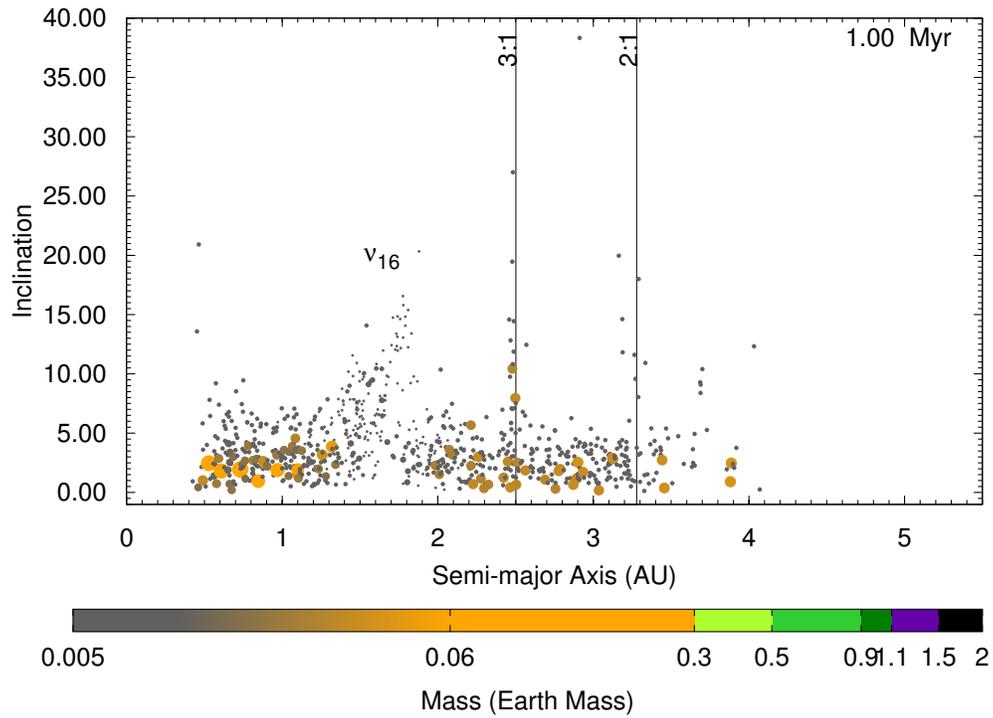}
\caption{Graph of the orbital inclination versus semimajor axis for the first 1 Myr of the simulation of Figure 3.
As shown here, $\nu_{16}$ and mean-motion resonances with Jupiter increase the inclinations of 
planetesimals and planetary embryos. This orbital excitation is more pronounced in the depleted region
(1.3-2.1 AU).}
\end{figure}

\clearpage
\begin{figure}
\centering
\includegraphics[scale=0.75]{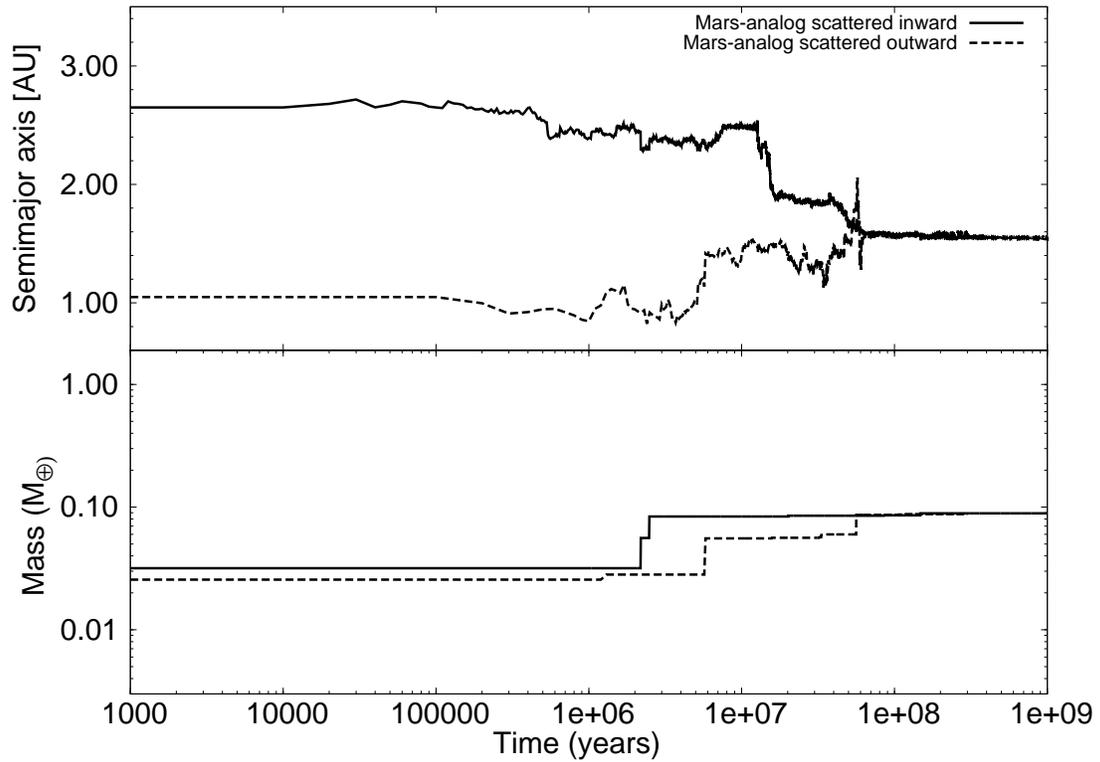}
\caption{Graphs of the semimajor axes and masses of two typical Mars-analogs formed in simulations where Jupiter and Saturn 
were initially in their current orbits. Similar to the results shown here, Mars-analogs in all our simulations were 
formed in the inner or outer non-depleted regions and were scattered into the depleted area.}
\end{figure}

\clearpage
\begin{figure}
\vspace{-1.7cm}
\includegraphics[scale=.75]{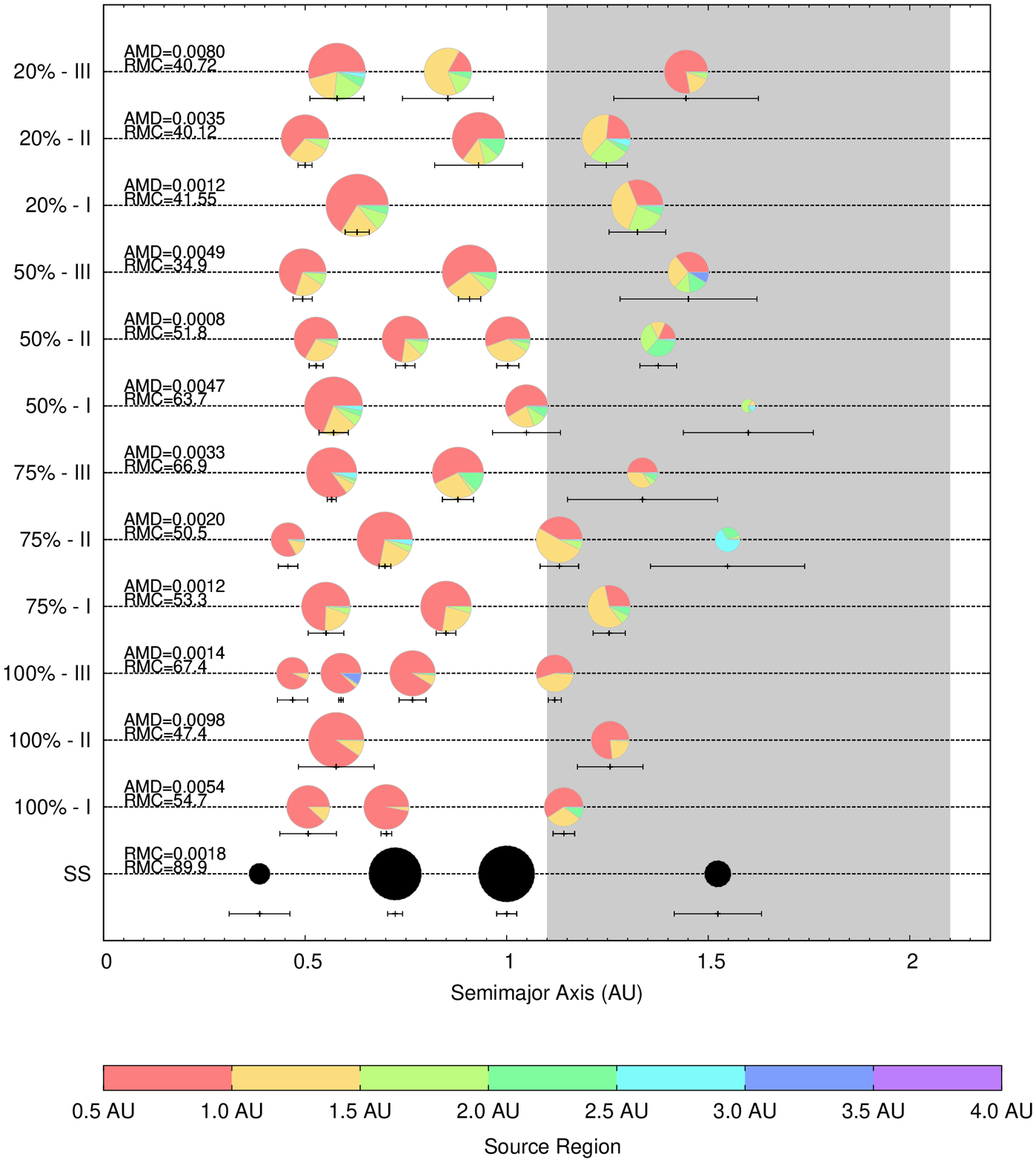}
\caption{Final masses and orbital configurations of planets in simulations of disk model A with different mass-depletion scales.
The depleted area is from 1.1 AU to 2.1 AU (shown in gray). As shown by the Roman numbers on the vertical axis, 
each simulation was run for three different distributions of planetesimals and planetary embryos. The size of each 
body corresponds to its relative physical size scaled as $M^{1/3}$. However, it is not to scale on the $x$-axis. 
The color of each object represents the relative contribution of material from different parts of the disk. 
The eccentricity of each planet is represented by its variation in heliocentric distance over 
the semimajor axis (horizontal bars).}
\end{figure}

\clearpage
\begin{figure}
\vspace{-1.7cm}
\centering
\includegraphics[scale=.75]{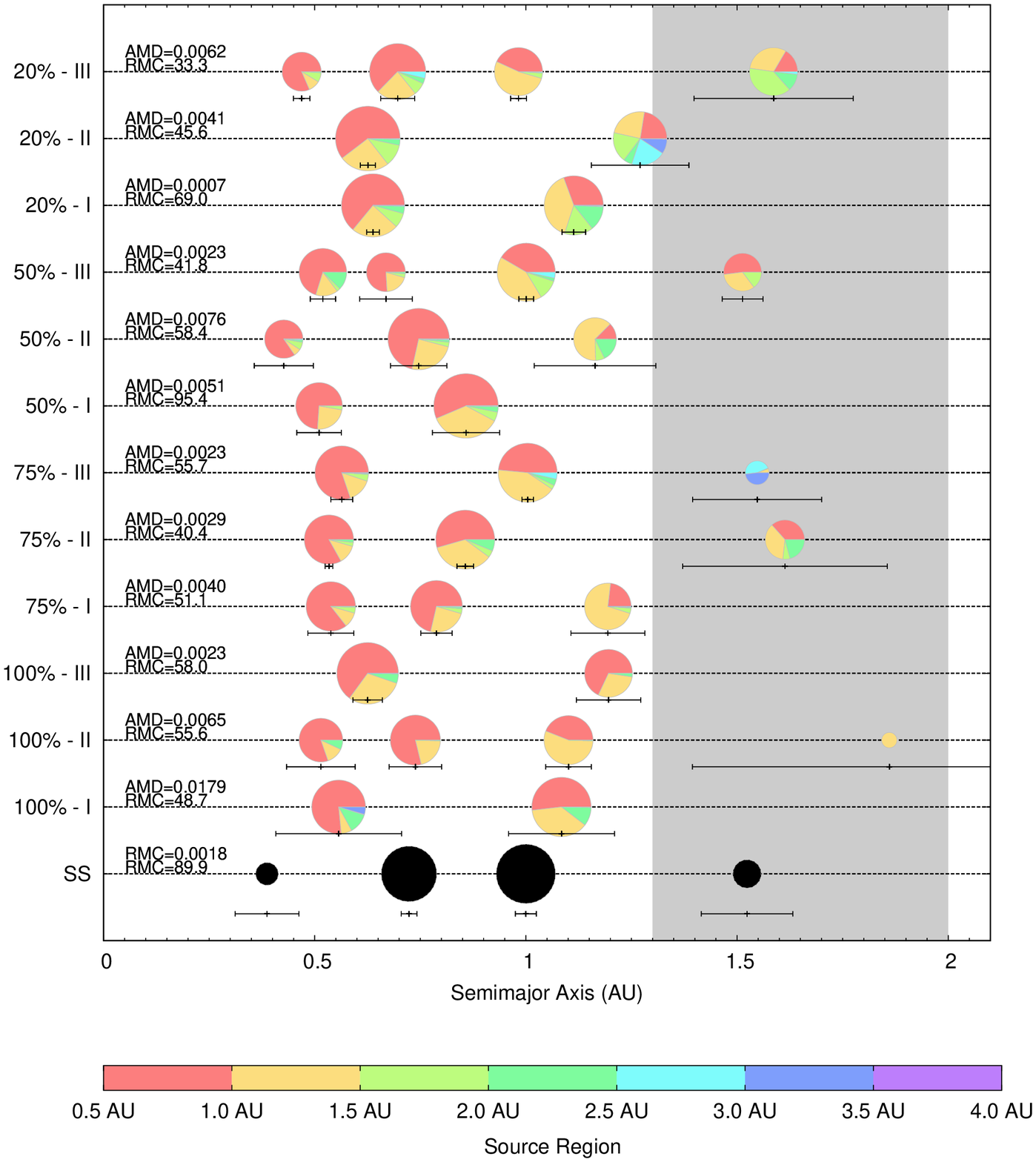}
\caption{Final masses and orbital configurations of planets in simulations of disk model B with different mass-depletion scales.
The depleted area is from 1.3 AU to 2.0 AU (shown in gray). As shown by the Roman numbers on the vertical axis, 
each simulation was run for three different distributions of planetesimals and planetary embryos. The size of each 
body corresponds to its relative physical size scaled as $M^{1/3}$. However, it is not to scale on the $x$-axis. 
The color of each object represents the relative contribution of material from different parts of the disk. 
The eccentricity of each planet is represented by its variation in heliocentric distance over 
the semimajor axis (horizontal bars).}
\end{figure}

\clearpage
\begin{figure}
\centering
\includegraphics[scale=0.7]{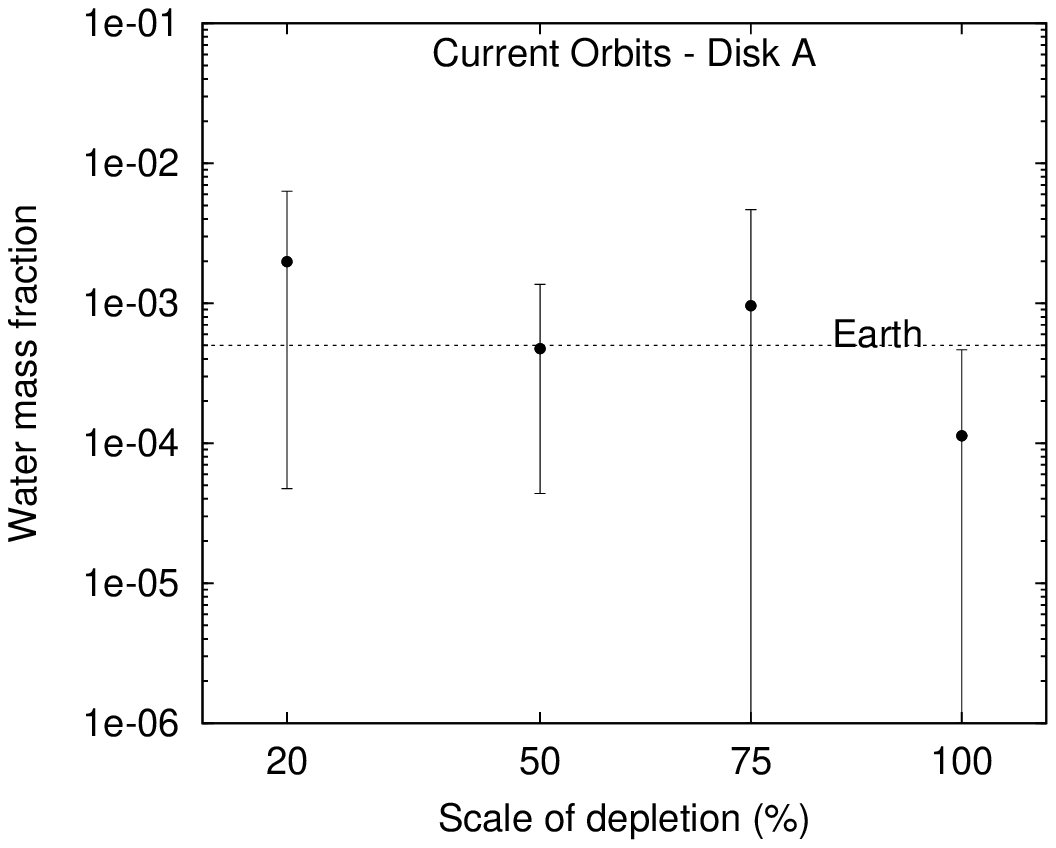}
\includegraphics[scale=0.7]{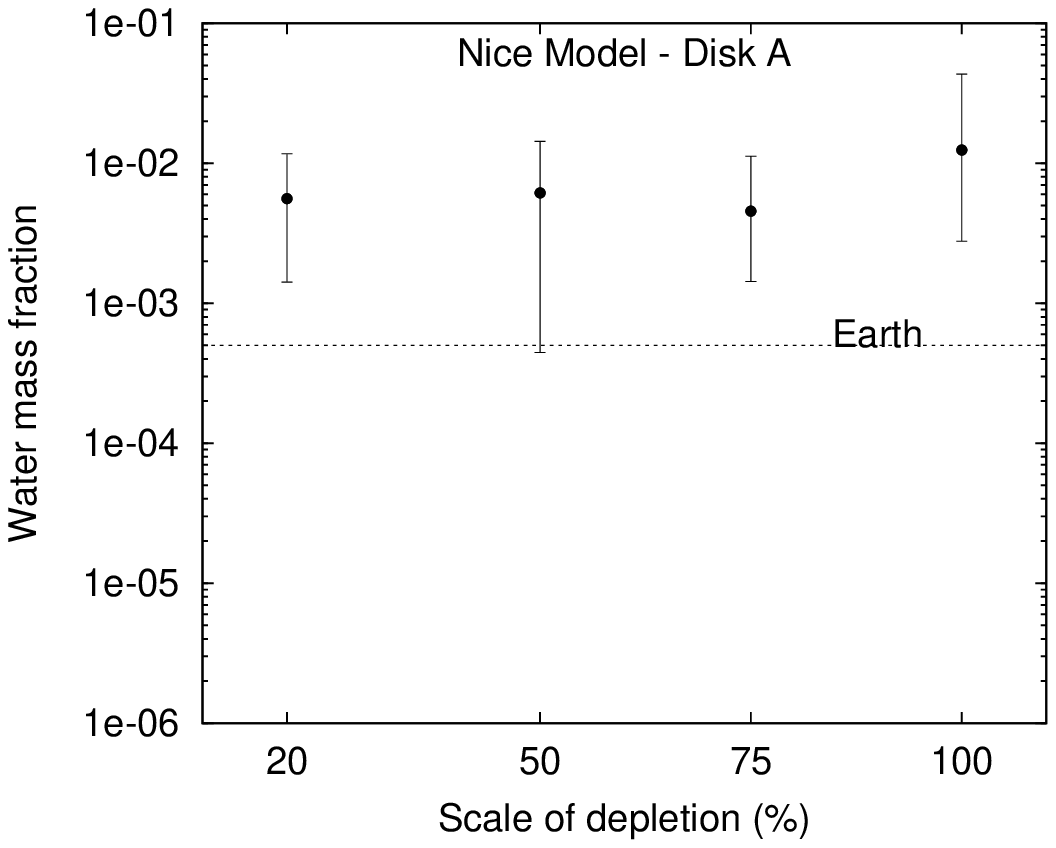}
\includegraphics[scale=0.7]{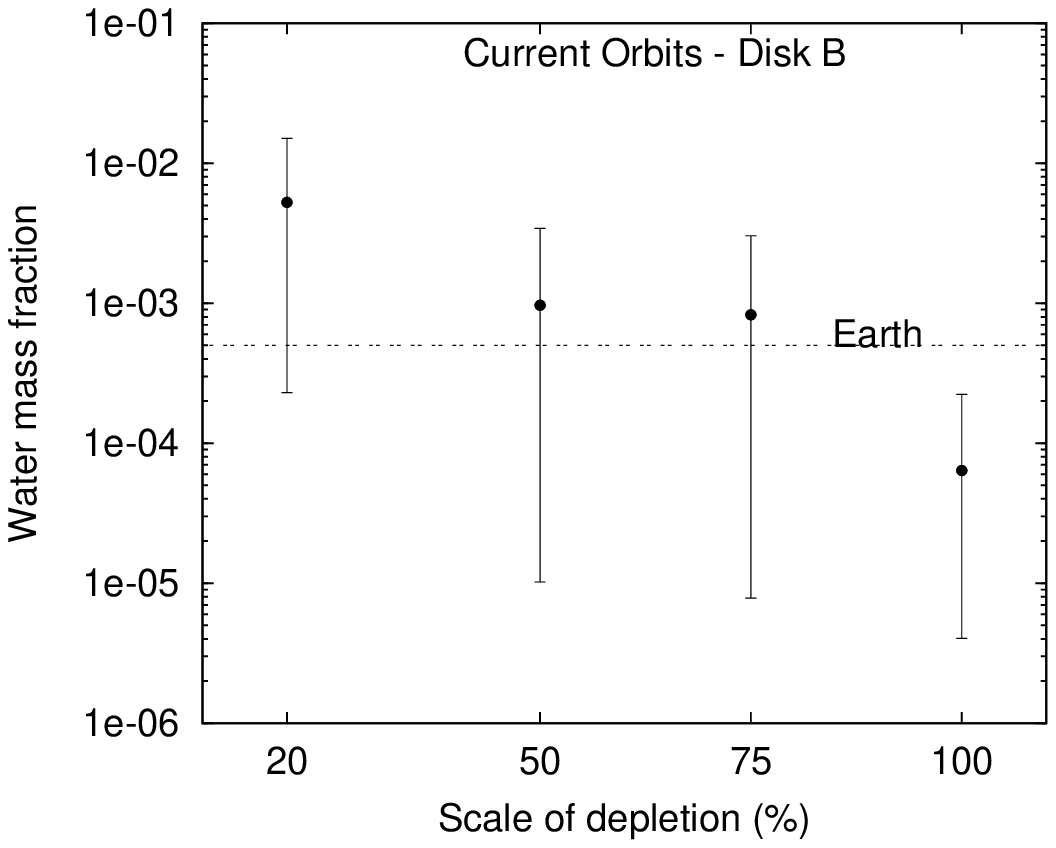}
\includegraphics[scale=0.7]{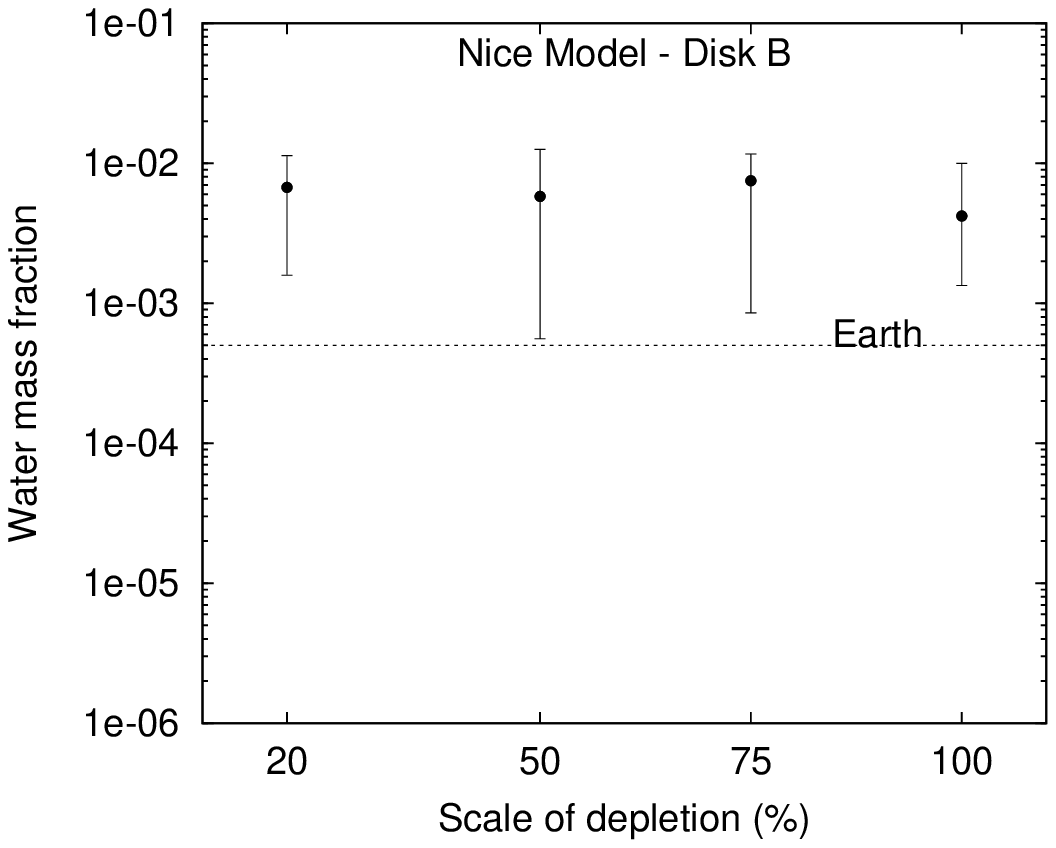}
\caption{Graphs of the mean value of the water/mass fraction of the planets formed inside 1.3 AU as a function of the scale of 
the disk mass-depletion for different disk models and giant planets configuration. The panels on the left show 
the results of the simulations considering Jupiter and Saturn to be initially in their current orbits. The ones on the right
show the results for which Jupiter and Saturn were initially as in the Nice Model (2005).}
\end{figure}

\clearpage
\begin{figure}
\centering
\includegraphics[scale=1]{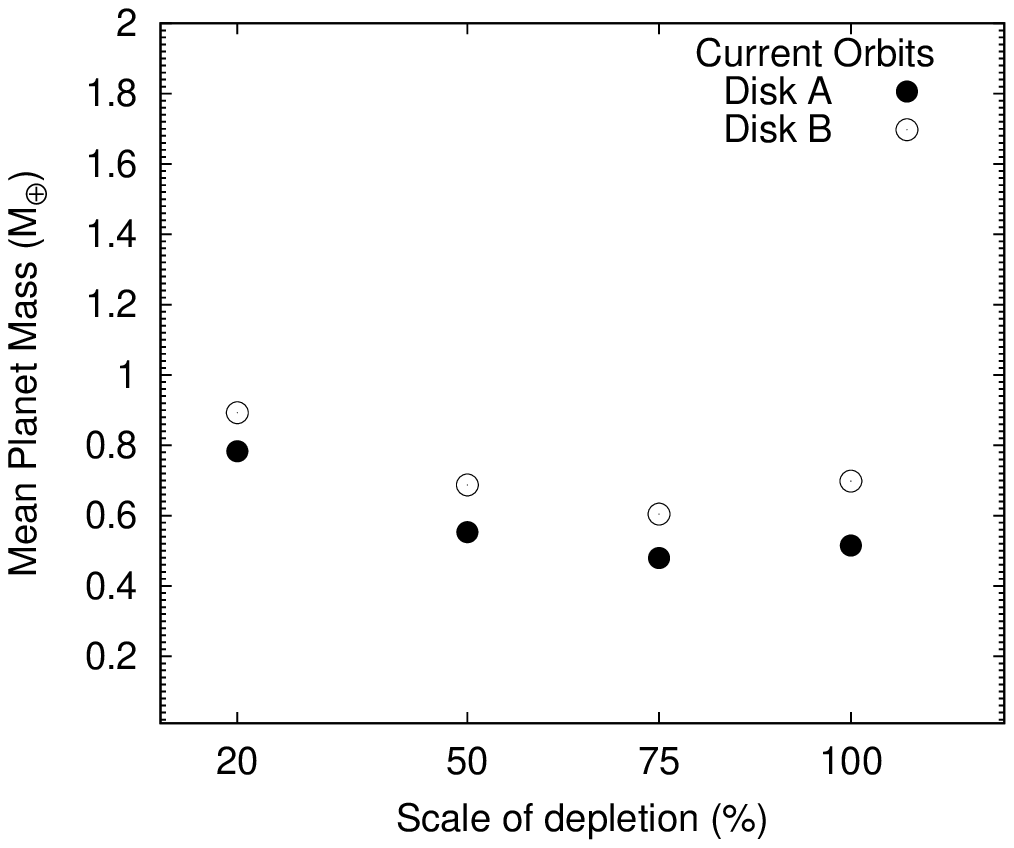}
\includegraphics[scale=1]{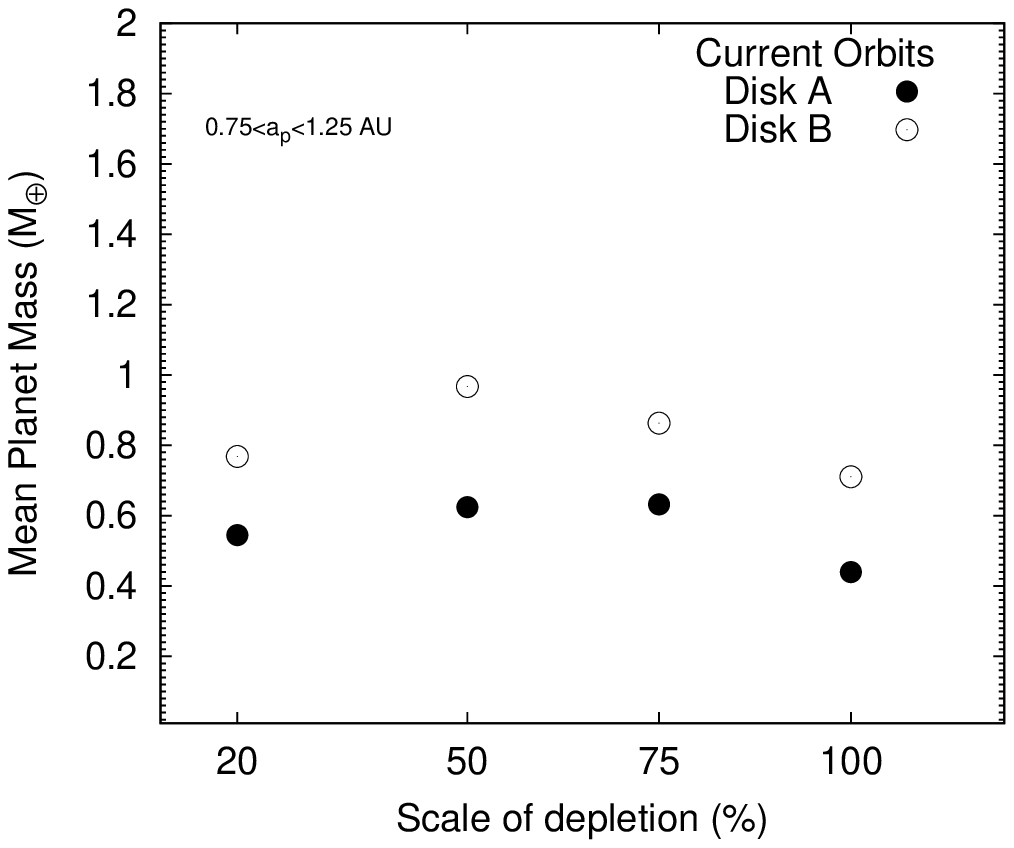}
\caption{Graphs of the mean value of planet mass as a function of the scale of mass-depletion in different disk models.
Jupiter and Saturn were initially in their current orbits. The top panel shows the mean planet mass for planets formed
interior to 2 AU and the bottom panel shows similar quantity for planets formed around 1 AU (0.75$< {a_p} <$ 1.25 AU).}
\end{figure}

\clearpage
\begin{figure}
\centering
\includegraphics[scale=0.75]{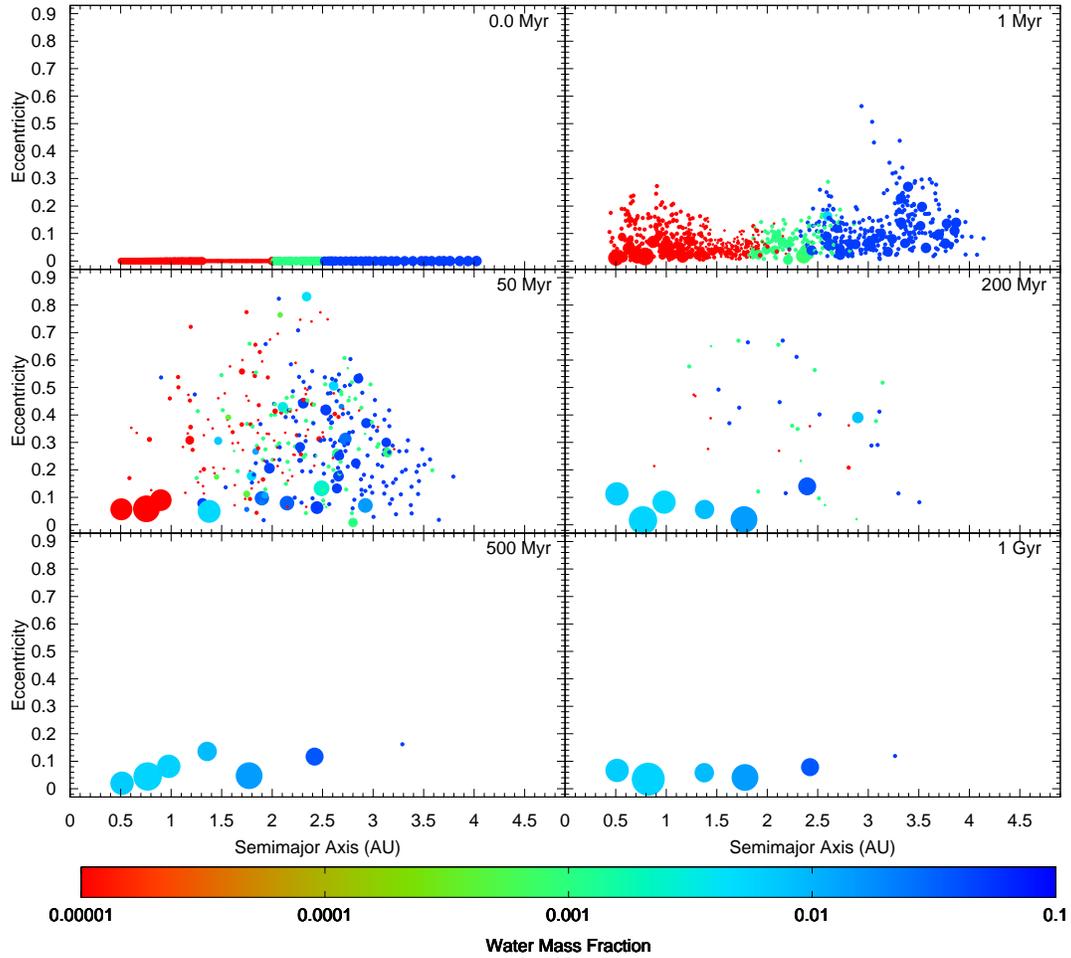}
\caption{Snapshots of the formation and dynamical evolution of planets in a disk with a depletion of 75\%  
extending from 1.3 AU to 2.0 AU. Jupiter and Saturn were initially in circular orbits corresponding to the Nice Model. 
The size of each body corresponds to its relative physical size and is scaled as $M^{1/3}$. However, it is not to 
scale on the $x$-axis. The color-coding represents the water-mass fraction of the body.}
\end{figure}

\clearpage
\begin{figure}
\centering
\includegraphics[scale=1]{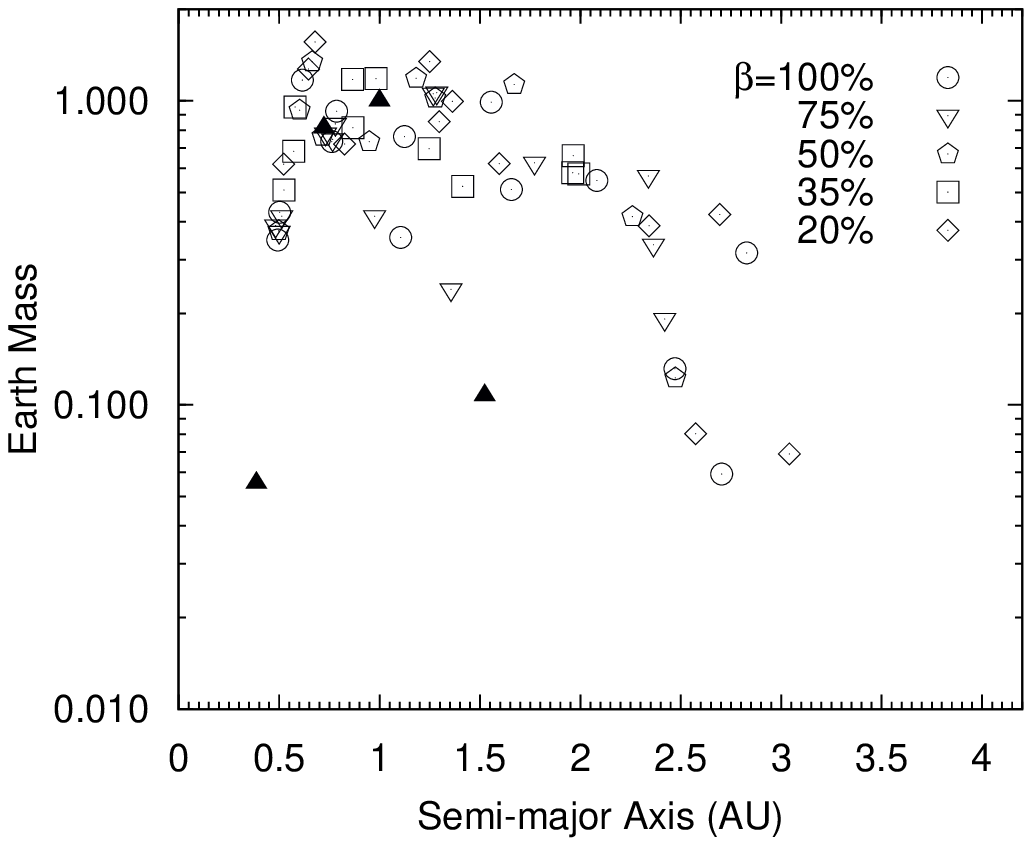}
\includegraphics[scale=1]{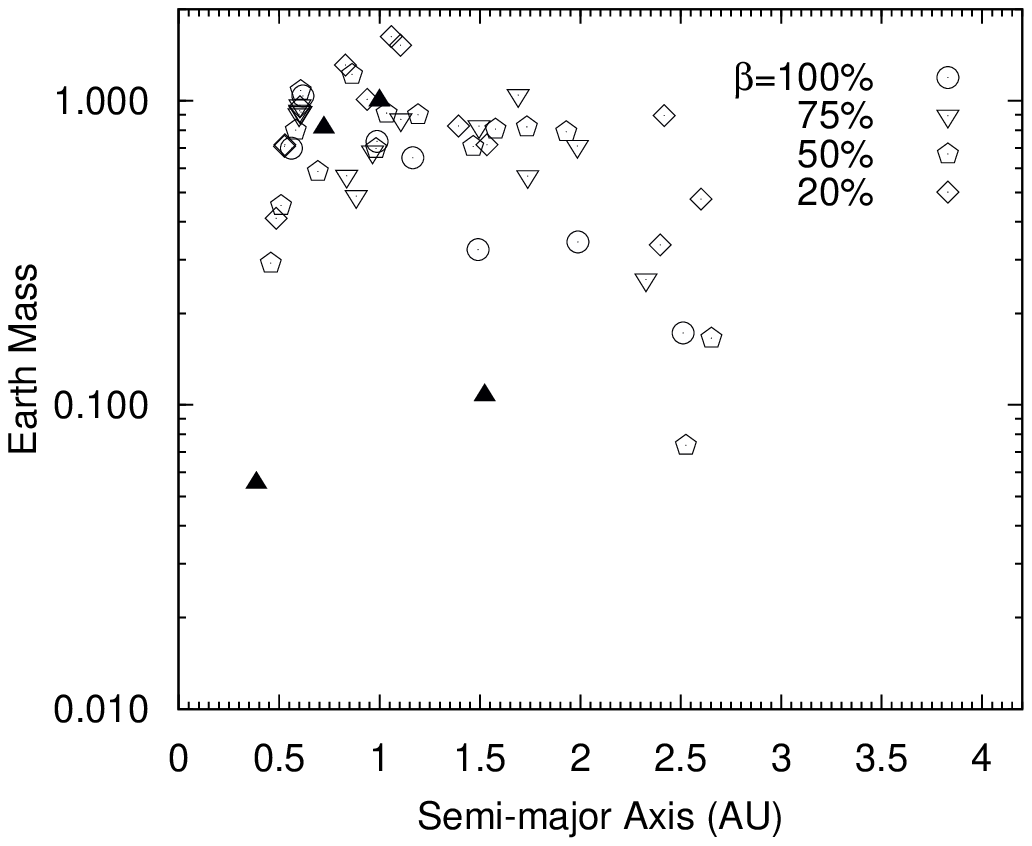}
\caption{Graphs of the mass-semimajor axis distribution for the surviving bodies (open circles) 
in the simulations considering the disk model A (top) and disk model B (bottom) for all scales of depletion. 
Jupiter and Saturn were initially in circular orbits. The solid triangles represent the masses of
Mars, Earth, Venus, and Mercury. The Surviving planetesimals are not shown.}
\end{figure}

\clearpage
\begin{figure}
\centering
\includegraphics[scale=1]{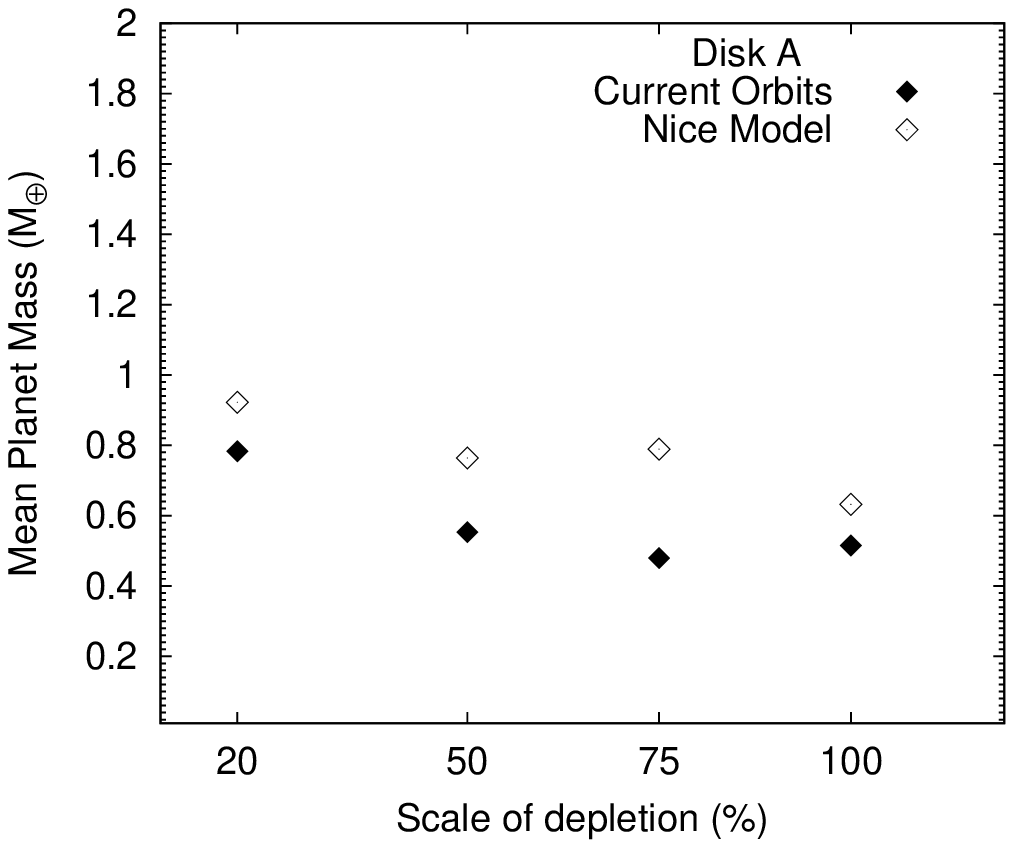}
\includegraphics[scale=1]{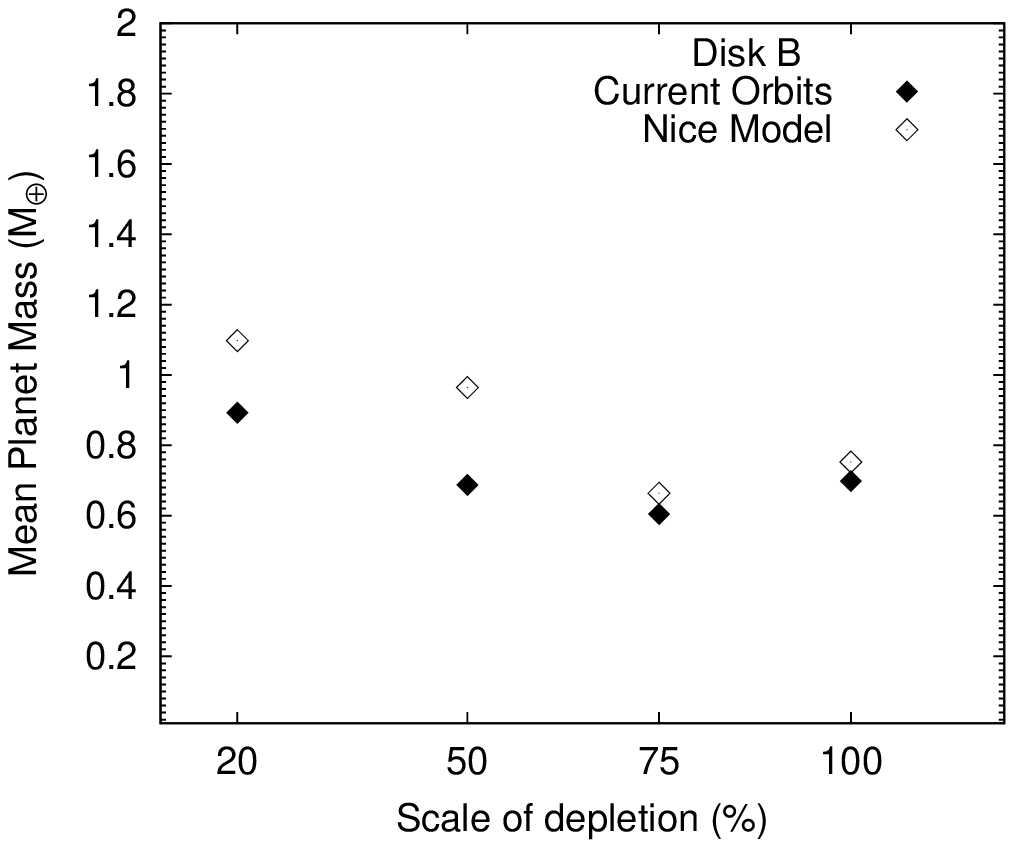}
\caption{Graphs of the mean values of the masses of planets in terms of the scale of mass-depletion 
for different disk models and giant planets configurations. The top panel corresponds to the disk model A and 
the bottom panel is for the disk model B.}
\end{figure}

\clearpage
\begin{figure}
\centering
\includegraphics[scale=0.7]{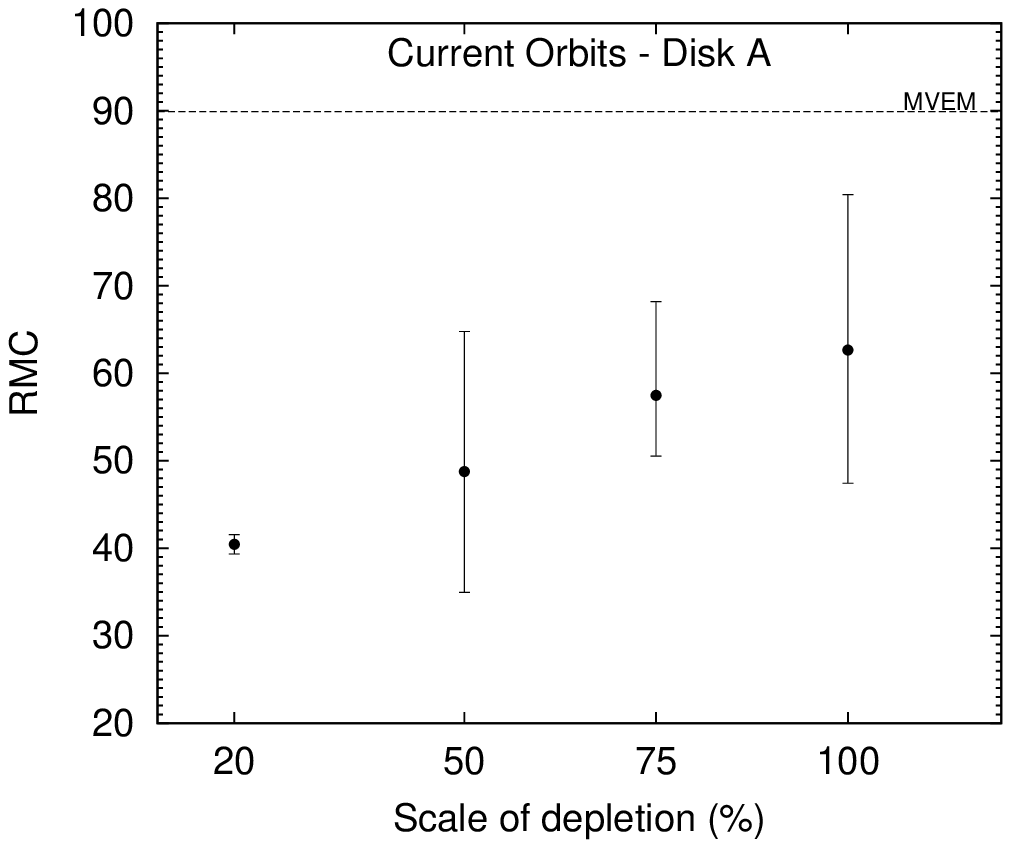}
\includegraphics[scale=0.7]{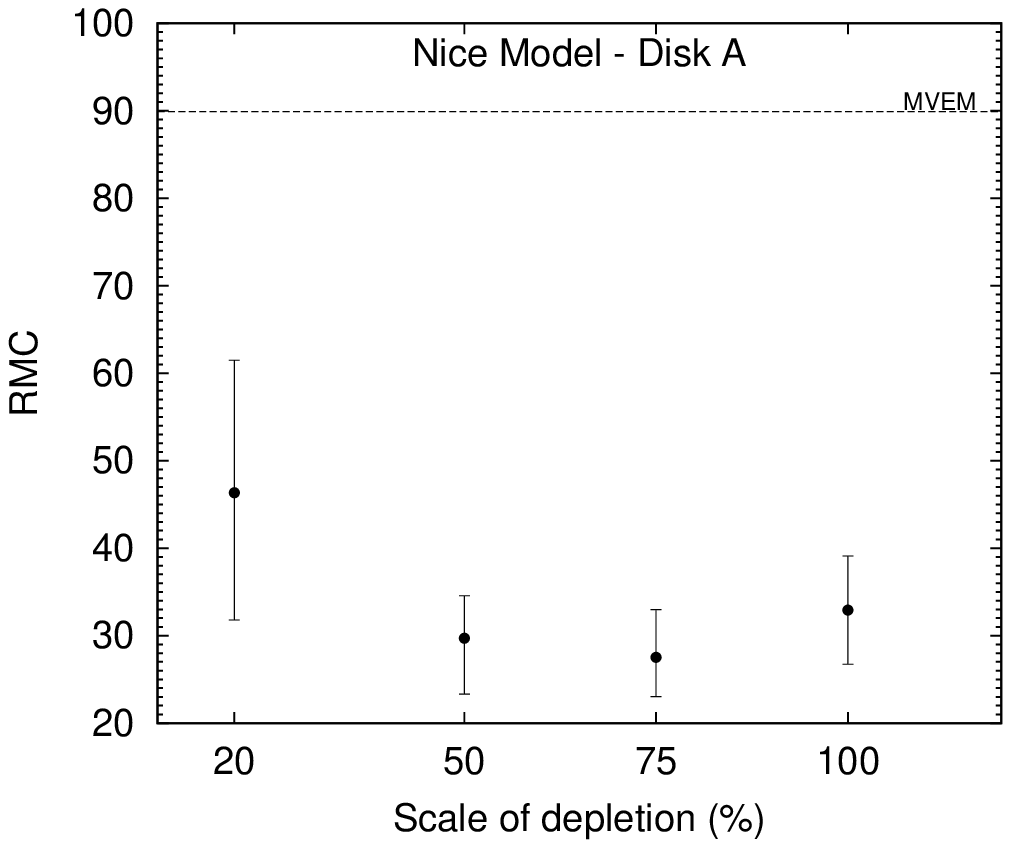}
\includegraphics[scale=0.7]{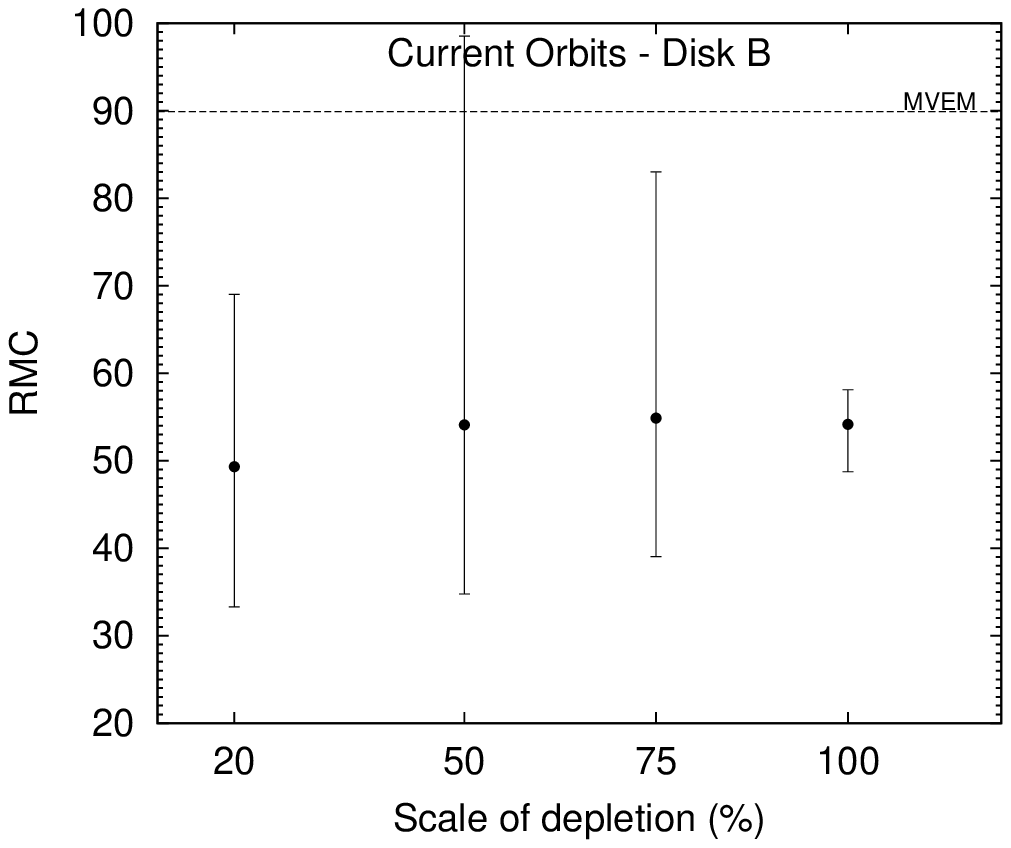}
\includegraphics[scale=0.7]{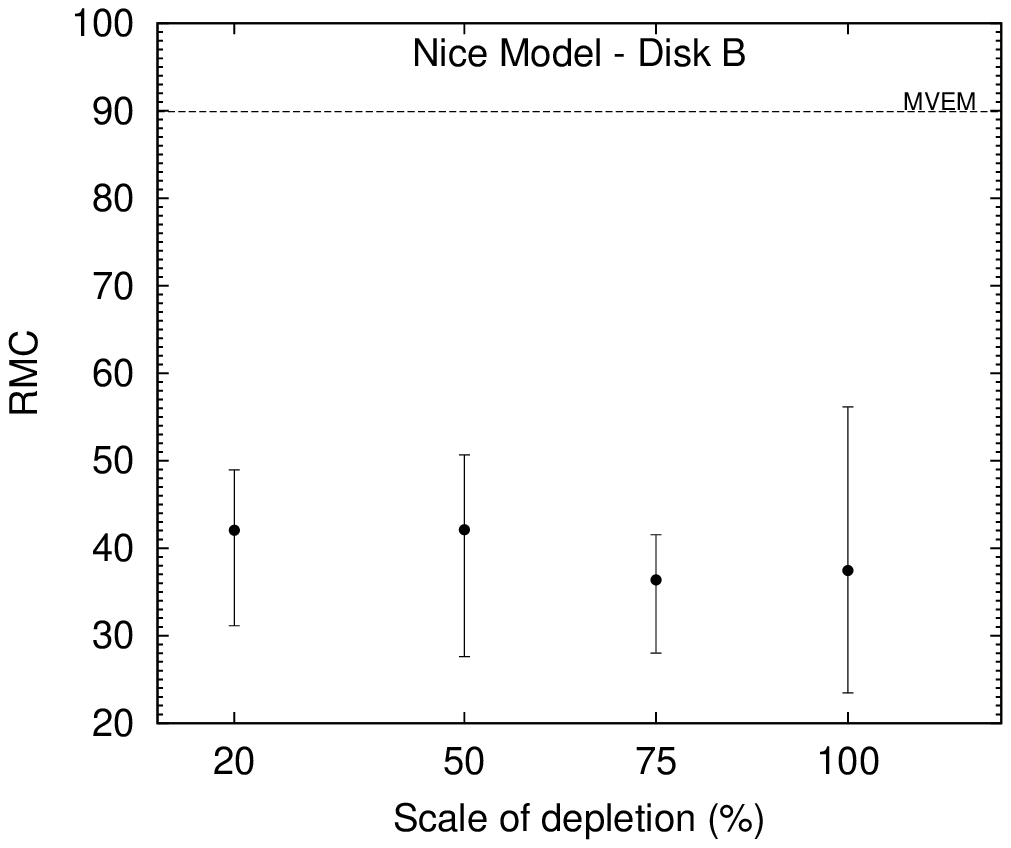}
\caption{Graphs of the mean values of RMC as a function of the mass-depletion scale for different disk models and 
giant planets configuration. The left column shows the results for
simulations in which Jupiter and Saturn were initially in their current orbits, and the right column shows the results
when these planets were in circular orbits similar to those in the Nice model (2005).}
\end{figure}

\clearpage
\begin{figure}
\centering
\includegraphics[scale=0.7]{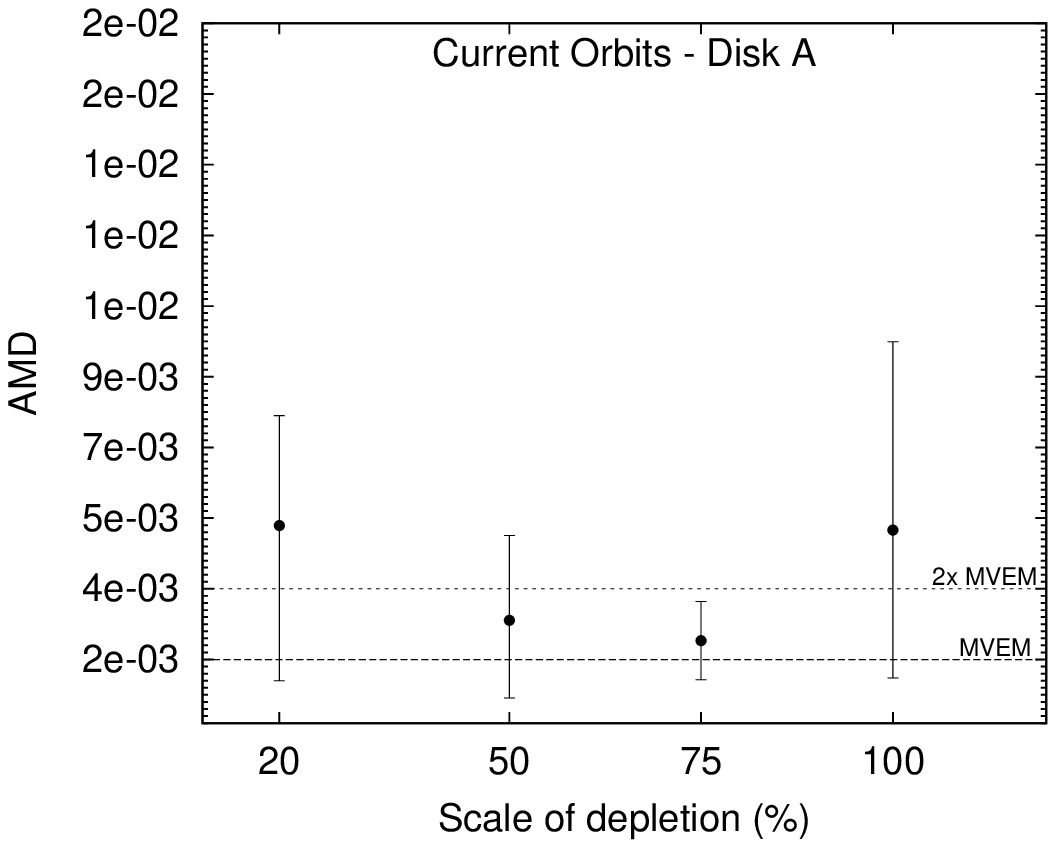}
\includegraphics[scale=0.7]{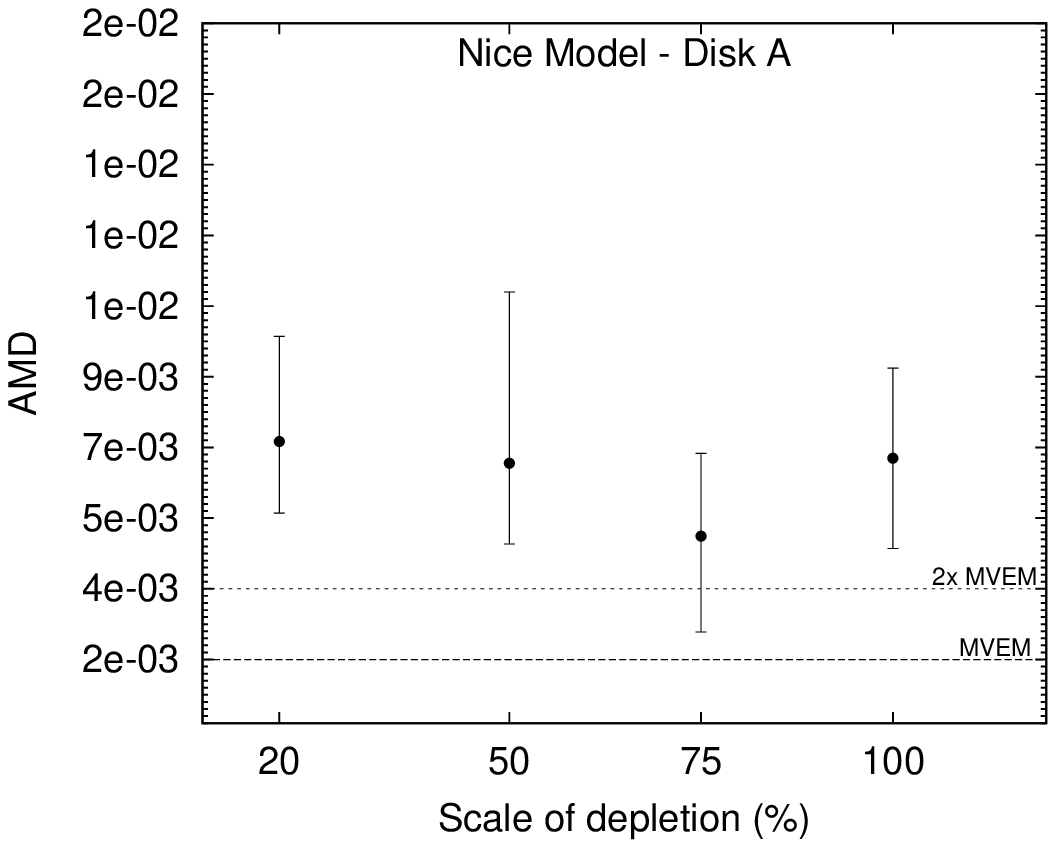}
\includegraphics[scale=0.7]{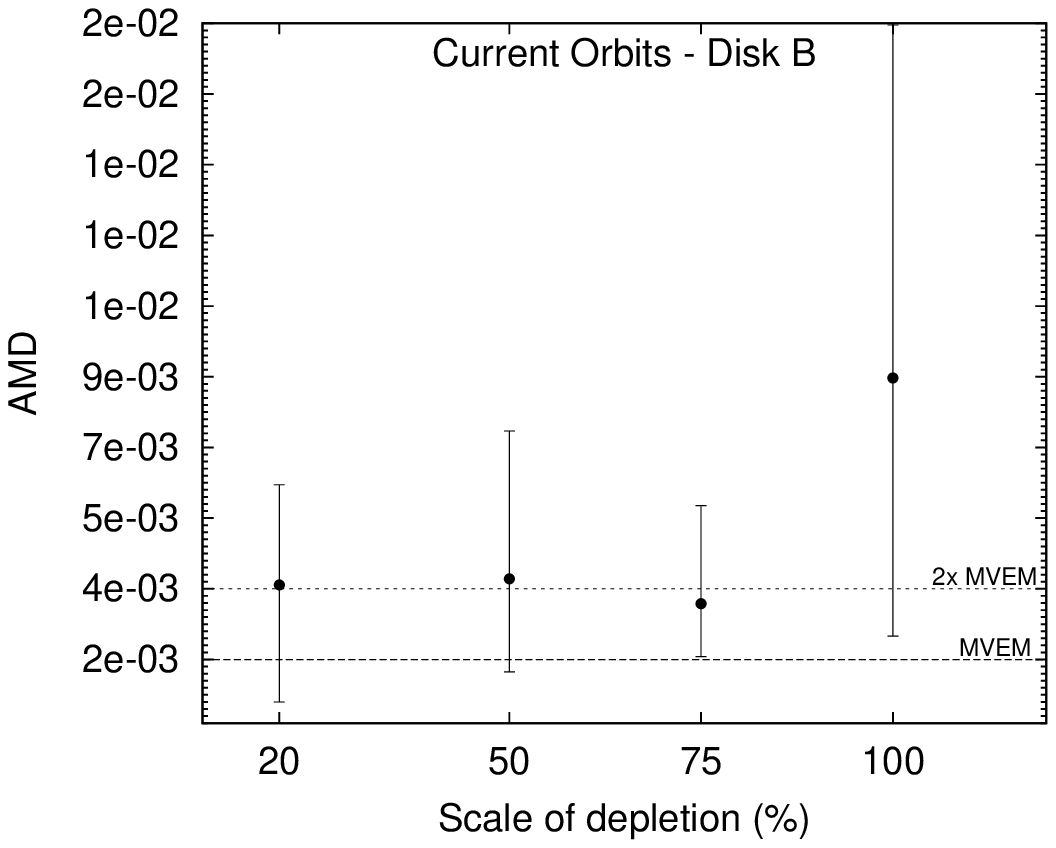}
\includegraphics[scale=0.7]{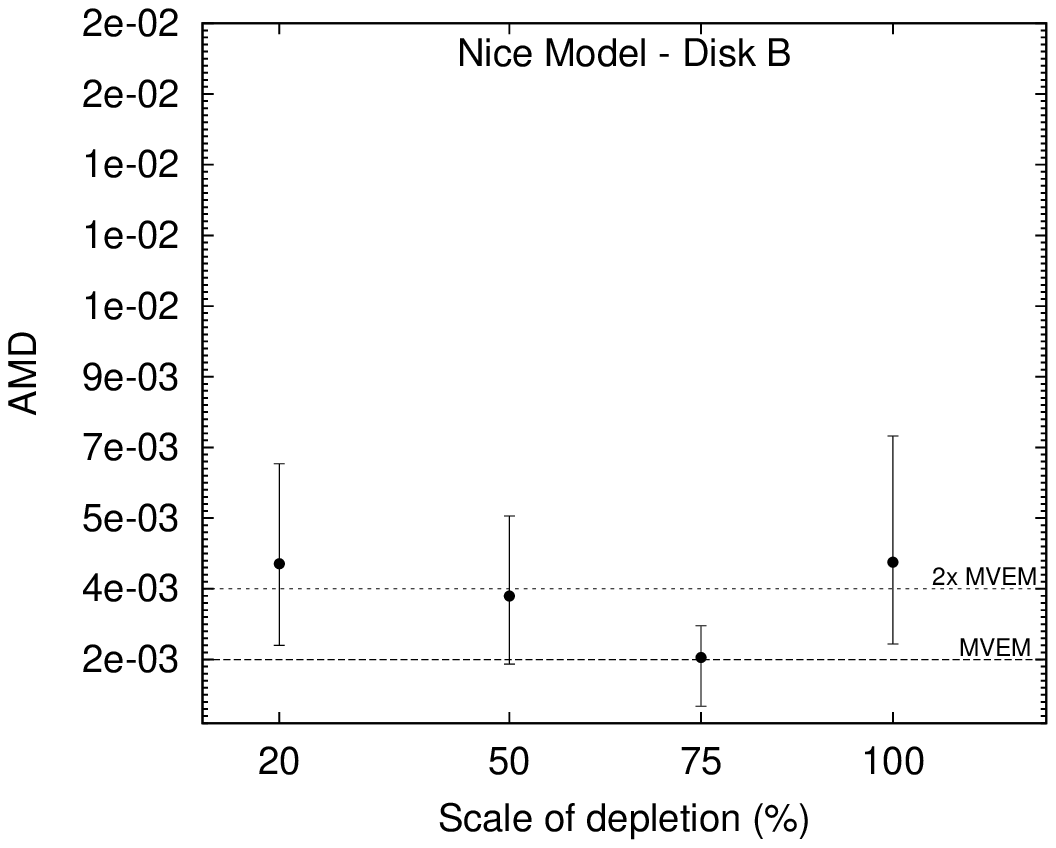}
\caption{Graphs of the mean values of AMD as a function of the mass-depletion scale for different disk models and 
giant planets configuration. The left column shows the results for
simulations in which Jupiter and Saturn were initially in their current orbits, and the right column shows the results
when these planets were in circular orbits similar to those in the Nice model (2005).}
\end{figure}

\clearpage
\begin{figure}
\centering
\includegraphics[scale=.7]{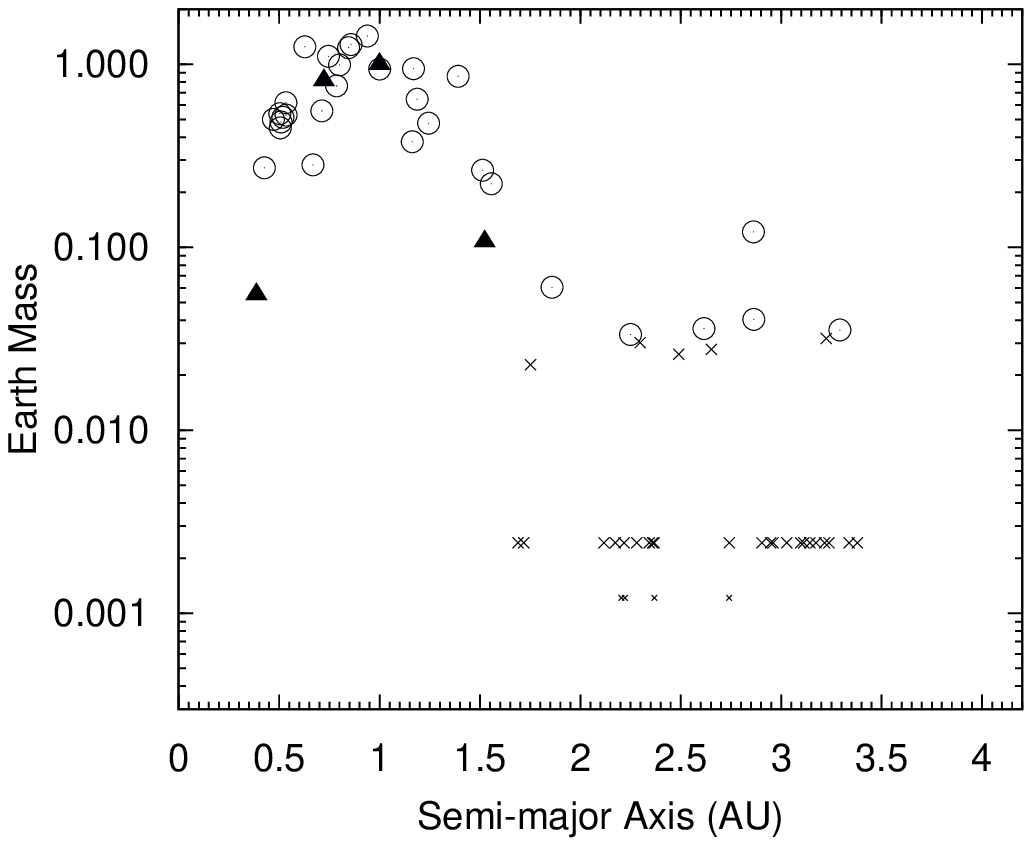}
\includegraphics[scale=.7]{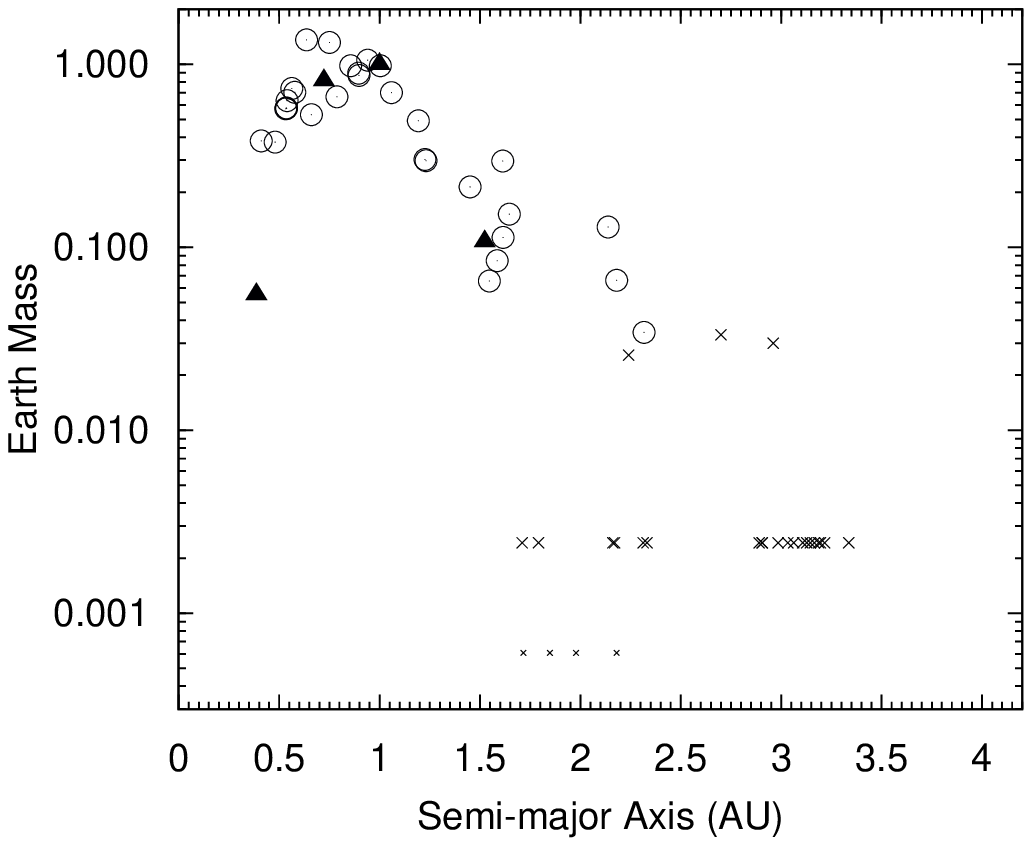}
\includegraphics[scale=.7]{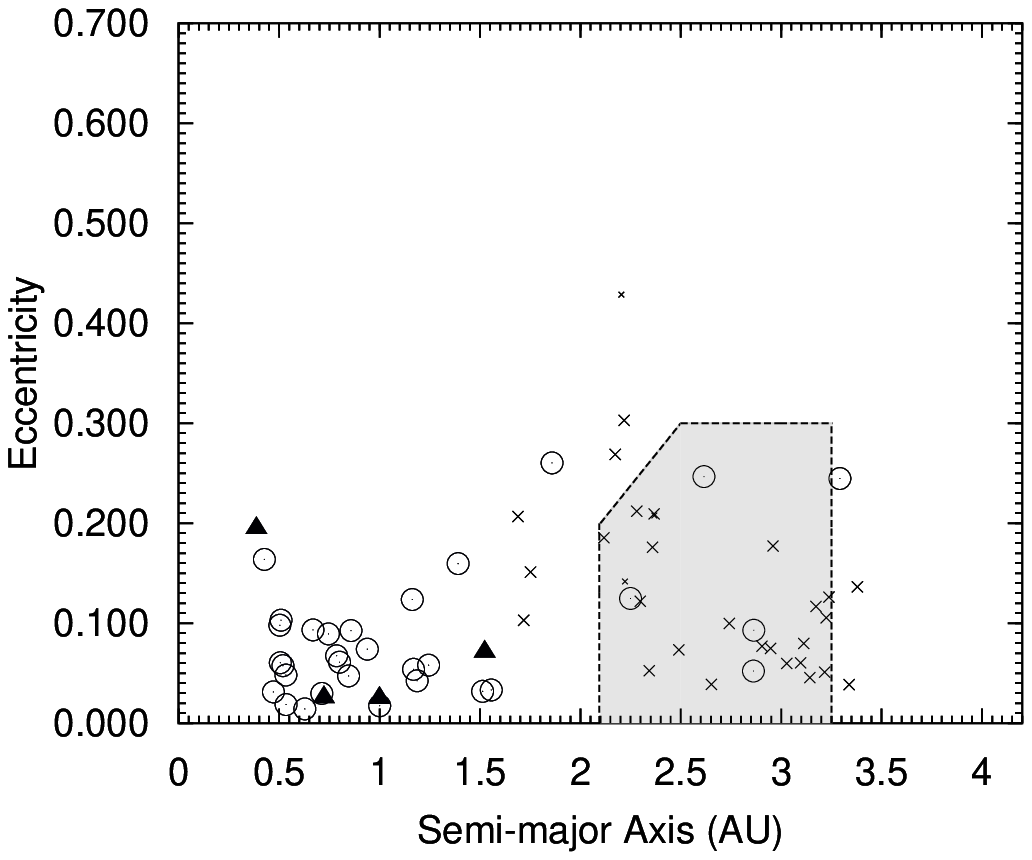}
\includegraphics[scale=.7]{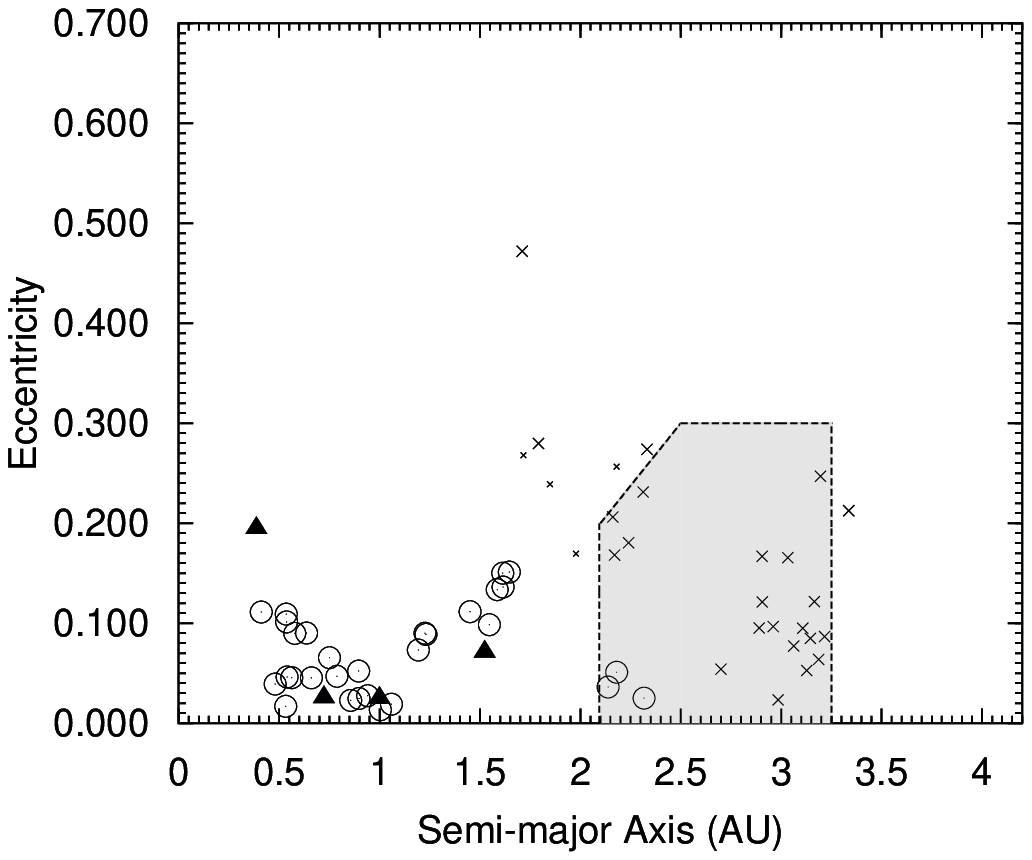}
\includegraphics[scale=.7]{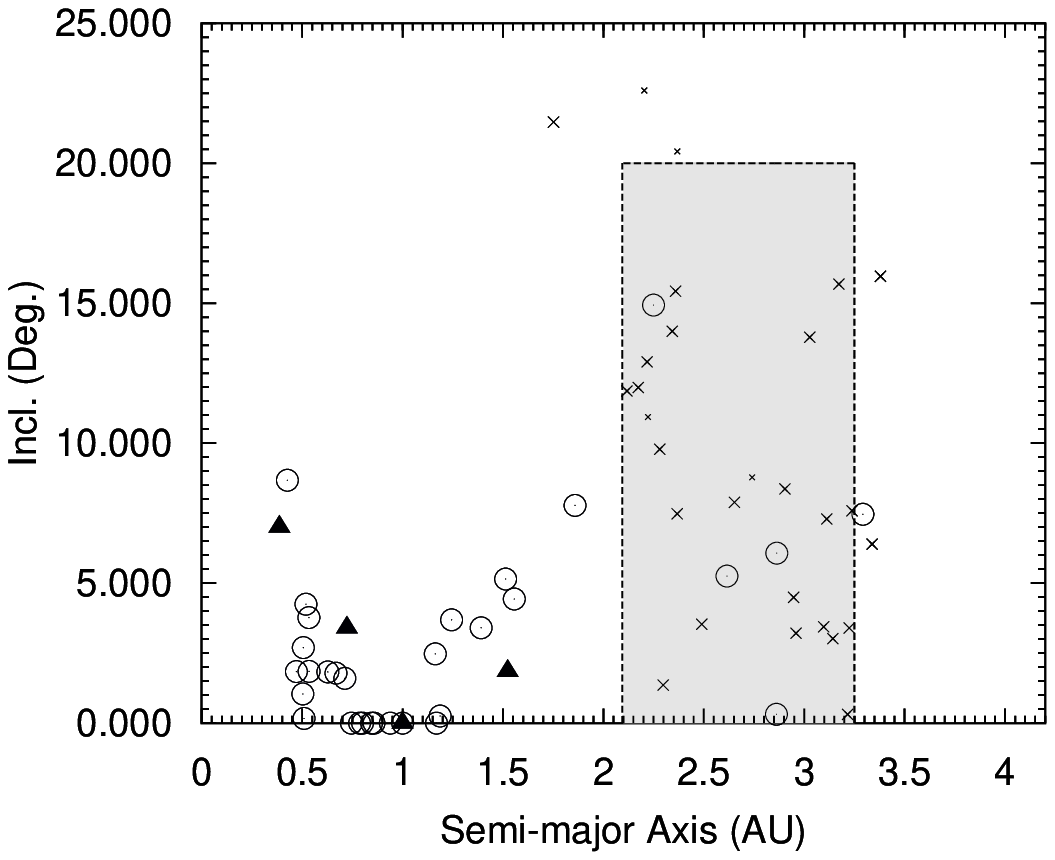}
\includegraphics[scale=.7]{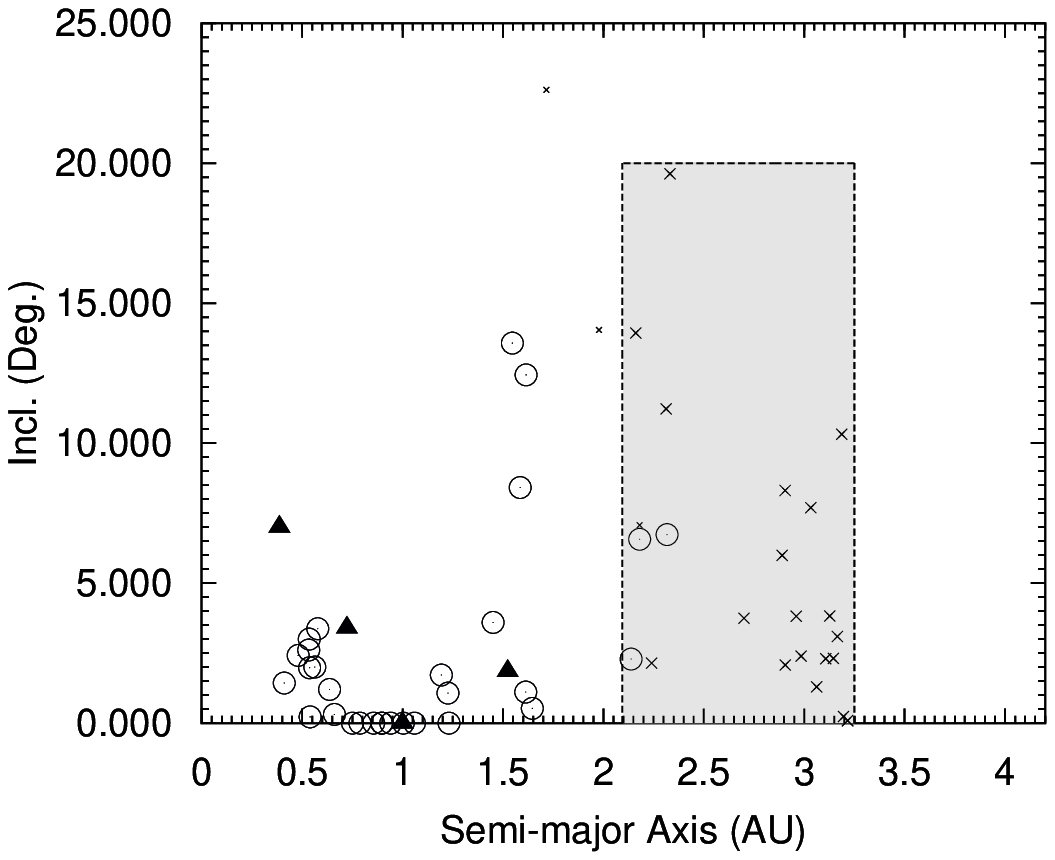}
\caption{Orbital distributions of the surviving bodies in the simulations considering the disk model B with Jupiter 
and Saturn initially in their current orbits. Open circles correspond to bodies with masses 
larger than $0.3\, M_{\rm Mars}\approx0.033 M_\oplus$. Smaller bodies are labeled with crosses. The left column
shows the results of 9 simulations  with a depletion scale of 50\% and the right column corresponds to those with 
a depletion scale of 75\%. The solid triangles represent the inner planets of the solar system. 
The gray area shows the asteroid belt.}
\end{figure}

\clearpage

\begin{table}
\centering
\begin{minipage}{80mm}
\caption{Regions and scales of mass-depletion}
\centering
\begin{tabular}{@{}lcl@{}}
  \hline
Disk & Region (AU) & \hskip 0.25in Scale (\%)  \\
  \hline\hline
A       & 1.1 to 2.1   & 20, 50, 75, 100   \\
\hline
B       & 1.3 to 2.0   & 20, 35, 50, 75, 100 \\
  \hline\hline
\end{tabular}
\end{minipage}
\end{table}

\clearpage

 \begin{table}
\centering
 \begin{minipage}{150mm}
 \scriptsize{
 \caption{Representative results of the simulations for the disk model A in which at least 
one body with a mass $M < 0.5M_{\oplus}$ was formed within 1.25 AU and 2.0 AU\tablenotemark{a}.}
\begin{tabular}{@{}llccccccllll@{}}
  \hline \hline
Sim  & Body  &   $a_{\rm init}$   & $a_{\rm fin}$  &  e &   I($^\circ$)    &  Mass &     WMF & $N_{c}$ &   $N_{gc}$   &   $t_{50\%}$   &      $t_{90\%}$  \\   
  &    &  (AU)   &  (AU) &    &      &  ($M_\oplus$) &    &      &  &   (Myr)   &  (Myr)    \\   \hline  \hline
A-100\%- II  & EM17  &   0.5988   & 0.5772  &   0.1626 &   11.9910    &  0.9479 &     2.581546e-06 &      43  &      10   &         114.90   &       562.30  \\ 
A-100\%- II  & EM36  &   0.7690   & 1.2567  &   0.0647 &   9.9490    &  0.3066 &     7.981451e-06 &      25  &       8   &           3.61   &       116.70  \\ 
A-100\%- II  & EM93  &   3.5587   & 2.6544  &   0.2764 &   4.0460    &  0.0667 &     5.000000e-02 &       1  &       1   &           3.40   &         3.40  \\ 
\hline
A-75\%- I   & EM5  &   0.5207   & 0.5516  &   0.0802 &   3.6940    &  0.6412 &     2.013308e-04 &      87  &      10   &          32.83   &       138.90  \\ 
A-75\%- I   & EM52  &   0.8923   & 0.8495  &   0.0288 &   2.9710    &  0.7274 &     0.00000e+00 &      91  &      12   &          39.44   &        53.38  \\ 
A-75\%- I   & EM64  &   1.0636   & 1.2543  &   0.0318 &   2.7140    &  0.4355 &     3.537550e-04 &      33  &       6   &          62.42   &       147.70  \\ 
A-75\%- I   & EM164  &   2.3522   & 2.2403  &   0.1213 &   3.2200    &  0.0256 &     1.000000e-03 &       0  &       0   &           0.00   &         0.00  \\ 
A-75\%- I   & EM173  &   2.7681   & 3.0647  &   0.0698 &   2.9860    &  0.0660 &     4.907275e-02 &       6  &       1   &           0.00   &         0.00  \\ 
\hline
A-75\%- II   & EM34  &   0.7457   & 0.4576  &   0.0528 &   4.2490    &  0.2195 &     5.723449e-04 &      40  &       6   &           4.05   &        40.76  \\ 
A-75\%- II   & EM16  &   0.5941   & 0.6979  &   0.0212 &   3.5730    &  0.9520 &     1.572659e-03 &      99  &      15   &          12.34   &        64.75  \\ 
A-75\%- II   & EM53  &   0.9574   & 1.1308  &   0.0423 &   3.5950    &  0.5470 &     4.593763e-04 &      63  &      13   &          12.92   &        28.36  \\ 
A-75\%- II   & EM167  &   2.6842   & 1.5480  &   0.1236 &   13.4810    &  0.0864 &     3.373369e-02 &       7  &       2   &           2.19   &         2.48  \\ 
A-75\%- II   & EM172  &   2.9346   & 2.8572  &   0.1416 &   6.4470    &  0.0645 &     4.952557e-02 &       4  &       1   &           0.60   &         0.60  \\ 
\hline
A-75\%- III   & EM12  &   0.5698   & 0.5661  &   0.0197 &   1.1750    &  0.7136 &     1.903639e-03 &      54  &      11   &          86.33   &       537.20  \\ 
A-75\%- III   & EM34  &   0.7334   & 0.8792  &   0.0439 &   6.1250    &  0.7475 &     1.227195e-04 &      77  &      14   &          12.94   &        21.26  \\ 
A-75\%- III   & EM45  &   0.8446   & 1.3370  &   0.1392 &   8.7110    &  0.1563 &     8.532762e-04 &      23  &       4   &          56.91   &        74.10  \\ 
A-75\%- III   & EM161  &   2.2421   & 2.2601  &   0.0261 &   0.5690    &  0.0578 &     9.788227e-04 &       6  &       1   &           0.00   &         0.00  \\ 
\hline
A-50\%- I   & EM19  &   0.6052   & 0.5708  &   0.0638 &   2.8780    &  1.0949 &     1.310486e-03 &      82  &      20   &          52.35   &       163.60  \\ 
A-50\%- I   & EM42  &   0.7977   & 1.0492  &   0.0803 &   3.7560    &  0.4318 &     6.430292e-04 &      41  &      11   &          10.94   &        71.66  \\ 
A-50\%- I   & EM113  &   1.7161   & 1.5994  &   0.1008 &   29.1090    &  0.0157 &     7.859689e-03 &       6  &       0   &           0.00   &        40.56  \\ 
A-50\%- I   & EM169  &   3.9663   & 3.9013  &   0.1441 &   1.9040    &  0.0379 &     5.000000e-02 &       0  &       0   &           0.00   &         0.00  \\ 
\hline
A-50\%- II   & EM4  &   0.5167   & 0.5274  &   0.0329 &   0.9110    &  0.4821 &     1.522684e-05 &      63  &       9   &          20.48   &        53.26  \\ 
A-50\%- II   & EM45  &   0.8718   & 0.7484  &   0.0320 &   3.6580    &  0.5544 &     2.354406e-04 &      80  &      13   &           3.17   &        54.09  \\ 
A-50\%- II   & EM47  &   0.8949   & 1.0026  &   0.0275 &   3.9390    &  0.5127 &     2.640177e-04 &      57  &      13   &           5.52   &        43.28  \\ 
A-50\%- II   & EM128  &   2.1000   & 1.3759  &   0.0332 &   5.4550    &  0.2402 &     1.363240e-03 &      19  &       7   &          28.94   &       216.20  \\ 
A-50\%- II   & EM148  &   3.0285   & 3.0188  &   0.1541 &   6.0760    &  0.0359 &     5.000000e-02 &       2  &       0   &           0.00   &         0.00  \\ 
\hline
A-50\%- IV   & EM18  &   0.6036   & 0.4939  &   0.0481 &   8.6880    &  0.5856 &     2.151852e-04 &      78  &      14   &           4.18   &        44.41  \\ 
A-50\%- IV   & EM16  &   0.5941   & 0.9080  &   0.0303 &   1.1200    &  0.9066 &     4.376598e-05 &      70  &      17   &          58.97   &        86.67  \\ 
A-50\%- IV   & EM133  &   2.3435   & 1.4509  &   0.1171 &   4.8840    &  0.3826 &     4.400611e-03 &      15  &       5   &          53.69   &       126.40  \\ 
\hline
\hline
\end{tabular}
\tablenotetext{a}{From left to right, the columns show the simulation, the body, 
the semimajor axis of body in the beginning of the simulation, its final semimajor axis, eccentricity, inclination (deg), 
water-mass fraction, total number of collisions, number of collisions with large objects (impactor mass $> 0.01M_\oplus$), 
time of reaching 50\% of the mass, and time of reaching 90\% of the mass, respectively.}
}
\end{minipage}
\end{table}

\clearpage%

\begin{table}
\centering

 \begin{minipage}{150mm}
 \scriptsize{
 \caption{ Representative results of the simulations for the disk model B in which at least 
one body with a mass $M < 0.5M_{\oplus}$ was formed within 1.25 AU and 2.0 AU\tablenotemark{a}.}
\begin{tabular}{@{}llccccccllll@{}}
  \hline \hline
Sim  & Body  &   $a_{init}$   & $a_{fin}$  &  e &   I($^\circ$)    &  Mass &     WMF & $N_{c}$ &   $N_{gc}$   &   $t_{50\%}$   &      $t_{90\%}$  \\   
  &    &  (AU)   &  (AU) &    &      &  ($M_\oplus$) &    &      &  &   (Myr)   &  (Myr)    \\   \hline  \hline
B-75\%- II   & EM26  &   0.6646   & 0.5338  &   0.0169 &   2.6090    &  0.5749 &     1.808761e-05 &      88  &      14   &           2.60   &        41.26  \\ 
B-75\%- II   & EM60  &   1.0287   & 0.8565  &   0.0229 &   5.2200    &  0.9882 &     6.242398e-05 &      61  &      14   &          72.66   &       149.60  \\ 
B-75\%- II   & EM70  &   1.1774   & 1.6135  &   0.1501 &   4.1080    &  0.2976 &     2.068806e-04 &      27  &       9   &          21.85   &       103.30  \\ 
\hline
B-75\%- III   & EM16  &   0.5973   & 0.5641  &   0.0457 &   5.3410    &  0.7419 &     1.715158e-04 &      58  &       8   &          54.60   &        93.92  \\ 
B-75\%- III   & EM5  &   0.5233   & 1.0044  &   0.0135 &   3.3400    &  0.9884 &     1.789666e-03 &      81  &      18   &           6.23   &        31.69  \\ 
B-75\%- III   & EM159  &   3.1186   & 1.5477  &   0.0987 &   16.9160    &  0.0658 &     4.674640e-02 &       3  &       1   &          18.16   &        85.59  \\ 
B-75\%- III   & EM137  &   2.1413   & 2.3173  &   0.0251 &   10.0770    &  0.0345 &     4.380566e-03 &       5  &       0   &           0.00   &         7.03  \\ 
\hline
B-50\%- II   & EM12  &   0.5742   & 0.4269  &   0.1639 &   10.8200    &  0.2736 &     4.650020e-04 &      37  &       6   &          24.02   &        73.82  \\ 
B-50\%- II   & EM19  &   0.6214   & 0.7462  &   0.0894 &   2.1440    &  1.1127 &     2.315770e-04 &      93  &      17   &          19.00   &        94.39  \\ 
B-50\%- II   & EM64  &   1.0799   & 1.1638  &   0.1237 &   4.6200    &  0.3797 &     1.764175e-04 &      22  &       7   &          52.41   &        70.54  \\ 
B-50\%- II   & EM136  &   2.6272   & 2.6524  &   0.0389 &   10.0330    &  0.0279 &     5.000000e-02 &       0  &       0   &           0.00   &         0.00  \\ 
\hline
B-50\%- III  & EM22  &   0.6366   & 0.5194  &   0.0577 &   6.0580    &  0.5161 &     7.913091e-05 &      69  &      10   &           5.34   &        61.23  \\ 
B-50\%- III  & EM25  &   0.6628   & 0.6688  &   0.0933 &   3.5970    &  0.2841 &     1.291794e-05 &      46  &       7   &           3.90   &        45.56  \\ 
B-50\%- III  & EM34  &   0.7396   & 1.0009  &   0.0174 &   1.8090    &  0.9474 &     1.656217e-03 &      85  &      15   &          17.57   &       116.50  \\ 
B-50\%- III  & EM10  &   0.5609   & 1.5132  &   0.0319 &   6.9570    &  0.2653 &     0.000000e+00 &      27  &       7   &          28.67   &       229.90  \\ 
B-50\%- III  & EM127  &   2.4728   & 2.8621  &   0.0521 &   1.4900    &  0.1222 &     2.729102e-02 &       5  &       3   &           1.58   &        25.55  \\ 
\hline
B-35\%- I   & EM38  &   0.7734   & 0.5684  &   0.0218 &   1.2620    &  0.8535 &     3.620376e-04 &      88  &      18   &          11.66   &        50.99  \\ 
B-35\%- I   & EM11  &   0.5580   & 0.9886  &   0.0331 &   6.3010    &  0.9829 &     2.852670e-04 &      66  &      18   &          21.94   &        92.21  \\ 
B-35\%- I   & EM53  &   0.9462   & 1.6636  &   0.1205 &   5.6370    &  0.3405 &     5.966567e-04 &      24  &       6   &          25.49   &        52.22  \\ 
\hline
\hline
\end{tabular}
\tablenotetext{a}{From left to right, the columns show the simulation, the body, 
the semimajor axis of body in the beginning of the simulation, its final semimajor axis, eccentricity, inclination (deg), 
water-mass fraction, total number of collisions, number of collisions with large objects (impactor mass $> 0.01M_\oplus$), 
time of reaching 50\% of the mass, and time of reaching 90\% of the mass, respectively.}
}
\end{minipage}
\end{table}

\clearpage

\begin{table}
\centering
 \begin{minipage}{160mm}
 \caption{Summary of the results of the simulations in disk model A. The heading of each column represents the criterion
that was used for assessing the success of a simulation. A success is indicated by ($\checkmark$), a failure is shown by ($\times$), and 
a  ($\sim$) indicates a near successful case\tablenotemark{a}}
\begin{tabular}{@{}lcccccccc@{}}
  \hline
 Sim. & $M_{\rm Mars}$ & $t_{\rm form,Mars}$ & $M_{\rm Earth}$ & $t_{\rm form,Earth}$ & $WMF_{\rm Earth}$ & AMD & $M_{\rm ast}$ & $N_{\rm ast}$ \\ \hline
A-75\%- I &   $\times$    &   $\times$    &   \checkmark   &   \checkmark   &   $\times$    &   \checkmark   &   $\times$    &  0  \\ \hline
A-75\% - II &  \checkmark    &   \checkmark    &   $\sim$    &   $\sim$    &   $\sim$    &    \checkmark     &   $\times$    &   1  \\  \hline
A-75\% - III &  \checkmark    &   $\times$    &   \checkmark    &   $\times$    &   $\times$    &  \checkmark &   \checkmark    &   6  \\  \hline
A-50\%- I &  $\times$    &   $\times$    &   $\times$    &   $\times$    &   $\times$    &   $\times$    &   \checkmark   &   1  \\ \hline
A-50\% - II &  \checkmark    &   $\times$    &   $\times$    &   $\times$    &   $\times$    &   \checkmark    &   \checkmark    &   4  \\  \hline
A-50\%- III &  $\times$    &   $\times$    &   \checkmark   &   \checkmark   &   $\times$    &   $\times$    &   \checkmark   &   9  \\    \hline
  \hline
\end{tabular}
\tablenotetext{a}{From left to right the columns represent the simulation, mass of the Mars-analog candidate,  
timescale of the formation of the Mars-analog candidate,  mass of the Earth candidate,  timescale of the formation of 
the Earth candidate, water-mass fraction of the Earth candidate, angular momentum deficit of the system (AMD), mass in embryos 
stranded in the asteroid belt, and the number of asteroids left.}
\end{minipage}
\end{table}

\clearpage

\begin{table}
\centering
 \begin{minipage}{160mm}
 \caption{Summary of the results of the simulations in disk model B. The heading of each column represents the criterion
that was used for assessing the success of a simulation. A success is indicated by ($\checkmark$), a failure is shown by ($\times$), and 
a  ($\sim$) indicates a near successful case\tablenotemark{a}}
\begin{tabular}{@{}lcccccccc@{}}
  \hline
 Sim. & $M_{\rm Mars}$    & $t_{\rm form,Mars}$ & $M_{\rm Earth}$ & $t_{\rm form,Earth}$ & $WMF_{\rm Earth}$ & AMD & $M_{\rm ast}$ & $N_{\rm ast}$ \\
  \hline\hline
B-75\% - I  &  $\times$    &  $\times$    &   \checkmark    &   $\times$    &   $\times$    &   $\times$    &   \checkmark    &   2   \\ \hline  
B-75\% - II &  \checkmark    &   $\times$    &   \checkmark    &   \checkmark    &   $\times$    &   \checkmark &    \checkmark    &   0 \\   \hline 
B-75\% - III &  \checkmark    &   $\times$    &   \checkmark    &   $\checkmark$    &   \checkmark    &   \checkmark   &   \checkmark     &   3  \\  \hline 

B-75\% - IV   &  \checkmark    &   $\times$    &   \checkmark    &   \checkmark    &   \checkmark    &   \checkmark    &   \checkmark      &  1  \\ \hline
B-75\% - V   &  \checkmark    &   $\times$    &   \checkmark    &   \checkmark    &   \checkmark    &   $\times$    &   $\times$    &  1   \\ \hline
B-75\% - VI   &  \checkmark    &   $\times$    &   \checkmark    &   $\times$    &   $\times$    &   \checkmark    &   \checkmark    &   2    \\ \hline
B-75\% - VII   &  $\times$    &   $\times$    &   \checkmark    &   \checkmark    &   \checkmark    &   $\times$    &   \checkmark    &  10  \\  \hline
B-75\% - VIII   &  \checkmark    &   $\times$    &   \checkmark    &   $\times$    &   $\times$    &   \checkmark    &   $\times$    &   0   \\ \hline
B-75\% - IX  &  $\times$    &   $\times$    &   $\times$    &   $\times$    &   $\times$    &   $\times$    &   \checkmark    &   0   \\ \hline

B-50\%- I &  $\times$    &   $\times$    &    \checkmark    &   $\times$    &   $\times$    &   $\times$    &    \checkmark    &   4 \\   \hline
B-50\%- II &  $\times$    &   $\times$    &   $\times$    &   $\times$    &   $\times$    &   $\times$    &    \checkmark    &   1 \\   \hline

B-50\%- III &  \checkmark    &   $\times$    &   \checkmark    &   \checkmark    &   \checkmark    &  \checkmark &   $\times$    &   2  \\  \hline
B-50\% - IV   &  \checkmark    &   $\times$    &   \checkmark    &   \checkmark    &   \checkmark    &   \checkmark    &   \checkmark    &   3   \\ \hline
B-50\% - V &  $\times$    &   $\times$    &   $\times$    &   $\times$    &   $\times$    &   $\times$    &   \checkmark     &   2  \\  \hline
B-50\% - VI &  \checkmark    &   \checkmark    &   \checkmark    &   $\sim$    &   \checkmark    &   $\times$    &   \checkmark   &   2 \\   \hline
B-50\% - VII &  $\times$    &   $\times$    &   \checkmark    &   $\times$    &   $\times$    &   \checkmark    &   $\times$    &   8   \\ \hline
B-50\% - VIII &  \checkmark    &   $\times$    &   \checkmark    &   \checkmark    &   $\times$    &   \checkmark    &  \checkmark    &   0  \\  \hline
B-50\% - IX &  $\times$    &   $\times$    &   \checkmark    &   \checkmark    &   \checkmark    &   \checkmark    &   \checkmark    &   2   \\ \hline
  
\end{tabular}
\tablenotetext{b}{From left to right the columns represent the simulation, mass of the Mars-analog candidate,  
timescale of the formation of the Mars-analog candidate,  mass of the Earth candidate,  timescale of the formation of 
the Earth candidate, water-mass fraction of the Earth candidate, angular momentum deficit of the system (AMD), mass in embryos 
stranded in the asteroid belt, and the number of asteroids left.}
\end{minipage}
\end{table}


\begin{thebibliography}{}




\bibitem[Agnor et al.(1999)]{1999Icar..142..219A} Agnor, C.~B., Canup, 
R.~M., \& Levison, H.~F.\ 1999, \icarus, 142, 219 


\bibitem[Agnor 
\& Lin(2012)]{2012ApJ...745..143A} Agnor, C.~B., \& Lin, D.~N.~C.\ 2012, \apj, 745, 143 
 

\bibitem[All{\`e}gre et al.(1995)]{1995GeCoA..59.1445A} All{\`e}gre, C.~J., Manh{\`e}s, G., Göupel, C.\ 1995, \gca, 59, 1445

\bibitem[Bouvier et al.(2007)]{2007GeCoA..71.1583B} Bouvier, A., Blichert-Toft, J., Moynier, F., Vervoort, J.~D., 
\& Albar{\`e}de, F.\ 2007, \gca, 71, 1583 

\bibitem[Brasser 
\& Morbidelli(2011)]{2011A&A...535A..41B} Brasser, R., \& Morbidelli, A.\ 2011, \aap, 535, A41 


\bibitem[Canup \& Pierazzo (2006)]{LPI...37...2146}
Canup, R. M. \& Pierazzo, E. \ 2006, LPI conference series, 37, 2146

\bibitem[Chambers(1998)]{1998EM&P...81....3C} Chambers, J.~E.\ 1998, Earth Moon and Planets, 81, 3 
\bibitem[Chambers(1999)]{1999MNRAS.304..793C} Chambers, J.~E.\ 1999, 
\mnras, 304, 793 
\bibitem[Chambers(2001)]{2001Icar..152..205C} Chambers, J.~E.\ 2001, 
\icarus, 152, 205 
\bibitem[Chambers(2007)]{2007Icar..189..386C} Chambers, J.~E.\ 2007, 
\icarus, 189, 386 

\bibitem[Chambers(2006)]{2006Icar..180..496C} Chambers, J.\ 2006, \icarus, 
180, 496 

\bibitem[Chambers(2013)]{2013Icar..224...43C} Chambers, J.~E.\ 2013, 
\icarus, 224, 43 


\bibitem[Chambers 
\& Cassen(2002)]{2002M&PS...37.1523C} Chambers, J.~E., \& Cassen, P.\ 2002, Meteoritics and Planetary Science, 37, 1523 
\bibitem[Chambers 
\& Wetherill(1998)]{1998Icar..136..304C} Chambers, J.~E., \& Wetherill, G.~W.\ 1998, \icarus, 136, 304 
\bibitem[Chambers 
\& Wetherill(2001)]{2001M&PS...36..381C} Chambers, J.~E., \& Wetherill, G.~W.\ 2001, 
Meteoritics and Planetary Science, 36, 381
\bibitem[Chiang 
\& Goldreich(1997)]{1997ApJ...490..368C} Chiang, E.~I., \& Goldreich, P.\ 1997, \apj, 490, 368 

\bibitem[Craddock 
\& Howard(2002)]{2002JGRE..107.5111C} Craddock, R.~A., \& Howard, A.~D.\ 2002, Journal of Geophysical Research (Planets), 107, 5111 

\bibitem[D'Angelo 
\& Marzari(2012)]{2012ApJ...757...50D} D'Angelo, G., \& Marzari, F.\ 2012, \apj, 757, 50 

\bibitem[Dauphas 
\& Pourmand(2011)]{2011Natur.473..489D} Dauphas, N., \& Pourmand, A.\ 2011, \nat, 473, 489 

\bibitem[Drouart et al.(1999)]{1999Icar..140..129D} Drouart, A., Dubrulle, 
B., Gautier, D., \& Robert, F.\ 1999, \icarus, 140, 129

\bibitem[Eisner et al.(2005)]{2005ApJ...623..952E} Eisner, J.~A., 
Hillenbrand, L.~A., White, R.~J., Akeson, R.~L., 
\& Sargent, A.~I.\ 2005, \apj, 623, 952 



\bibitem[Garaud 
\& Lin(2007)]{2007ApJ...654..606G} Garaud, P., \& Lin, D.~N.~C.\ 2007, \apj, 654, 606 

\bibitem[Genda \& Abe (2005)]{2005Nature...433...842M}
Genda, H. \& Abe, Y. \ 2005, Nature, 433, 842

\bibitem[Gladman et al.(1997)]{1997Sci...277..197G} Gladman, B.~J., 
Migliorini, F., Morbidelli, A., et al.\ 1997, Science, 277, 197 

\bibitem[Goldreich et al.(2004)]{2004ApJ...614..497G} Goldreich, P., 
Lithwick, Y., \& Sari, R.\ 2004, \apj, 614, 497 

\bibitem[Gomes(1997)]{1997AJ....114..396G} Gomes, R.~S.\ 1997, \aj, 114, 
396 

\bibitem[Gomes et al.(2005)]{2005Natur.435..466G} Gomes, R., Levison, 
H.~F., Tsiganis, K., \& Morbidelli, A.\ 2005, \nat, 435, 466 



\bibitem[Haghighipour et al.(2005)]{2012CMDA.....}  Haghighipour  N., Izidoro A., Winter O. C. 2013, submitted 
\bibitem[Hansen(2009)]{2009ApJ...703.1131H} Hansen, B.~M.~S.\ 2009, \apj, 
703, 1131 



\bibitem[Horner et 
al.(2007)]{2007EM&P..100...43H} Horner, J., Mousis, O., \& Hersant, F.\ 2007, Earth Moon and Planets, 100, 43 


\bibitem[Izidoro et al.(2013)]{2013ApJ...767...54I} Izidoro, A., de Souza 
Torres, K., Winter, O.~C., \& Haghighipour, N.\ 2013, \apj, 767, 54 



\bibitem[Jacobsen(2005)]{2005AREPS..33..531J} Jacobsen, S.~B.\ 2005, Annual 
Review of Earth and Planetary Sciences, 33, 531 


\bibitem[Jin et al.(2008)]{2008ApJ...674L.105J} Jin, L., Arnett, W.~D., 
Sui, N., \& Wang, X.\ 2008, \apjl, 674, L105 

\bibitem[Kobayashi 
\& Dauphas(2013)]{2013Icar..225..122K} Kobayashi, H., \& Dauphas, N.\ 2013, \icarus, 225, 122 


\bibitem[Kokubo 
\& Genda(2010)]{2010ApJ...714L..21K} Kokubo, E., \& Genda, H.\ 2010, \apjl, 714, L21 

\bibitem[Kokubo 
\& Ida(1998)]{1998Icar..131..171K} Kokubo, E., \& Ida, S.\ 1998, \icarus, 131, 171

\bibitem[Kokubo 
\& Ida(2000)]{2000Icar..143...15K} Kokubo, E., \& Ida, S.\ 2000, \icarus, 143, 15 

\bibitem[Kokubo et al.(2006)]{2006ApJ...642.1131K} Kokubo, E., Kominami, 
J., \& Ida, S.\ 2006, \apj, 642, 1131 

\bibitem[Laughlin et al.(2004)]{2004ApJ...608..489L} Laughlin, G., 
Steinacker, A., \& Adams, F.~C.\ 2004, \apj, 608, 489 
\bibitem[Laskar(1997)]{1997A&A...317L..75L} Laskar, J.\ 1997, \aap, 317, L75 

\bibitem[Lega et al.(2013)]{2013MNRAS.431.3494L} Lega, E., Morbidelli, A., 
\& Nesvorn{\'y}, D.\ 2013, \mnras, 431, 3494 

\bibitem[Levison 
\& Agnor(2003)]{2003AJ....125.2692L} Levison, H.~F., \& Agnor, C.\ 2003, \aj, 125, 2692 

\bibitem[Levison et al.(2011)]{2011AJ....142..152L} Levison, H.~F., 
Morbidelli, A., Tsiganis, K., Nesvorn{\'y}, D., 
\& Gomes, R.\ 2011, \aj, 142, 152 

\bibitem[Sofia Lykawka 
\& Ito(2013)]{2013arXiv1306.3287S} Sofia Lykawka, P., \& Ito, T.\ 2013, arXiv:1306.3287 

\bibitem[Lin 
\& Papaloizou(1986)]{1986ApJ...309..846L} Lin, D.~N.~C., \& Papaloizou, J.\ 1986, \apj, 309, 846 

\bibitem[Lunine et al.(2003)]{2003Icar..165....1L} Lunine, J.~I., Chambers, 
J., Morbidelli, A., \& Leshin, L.~A.\ 2003, \icarus, 165, 1 

\bibitem[Marty \& Yokochi (2006)]{2006Rev.Mineral.Geochem....62...421M}
Marty, B. \& Yokochi, R. \ 2006, Rev. mineral. Geochem., 62, 421

\bibitem[Marty(2012)]{2012E&PSL.313...56M} Marty, B.\ 2012, Earth and Planetary Science Letters, 313, 56 

\bibitem[Morbidelli et 
al.(2000)]{2000M&PS...35.1309M} Morbidelli, A., Chambers, J., Lunine, J.~I., et al.\ 2000, Meteoritics and Planetary Science, 35, 1309 

\bibitem[Morbidelli 
\& Crida(2007)]{2007Icar..191..158M} Morbidelli, A., \& Crida, A.\ 2007, \icarus, 191, 158 


\bibitem[Morbidelli et al.(2005)]{2005Natur.435..462M} Morbidelli, A., 
Levison, H.~F., Tsiganis, K., \& Gomes, R.\ 2005, \nat, 435, 462 

\bibitem[Morbidelli et al.(2012)]{2012AREPS} Morbidelli, A., Lunine, J.I., O'Brien, D.P., Raymond, S.N., 
Walsh, K.J. 2012, Annual Reviews of Earth and Planetary Science, in press.

\bibitem[Morishima et 
al.(2013)]{2013E&PSL.366....6M} Morishima, R., Golabek, G.~J., \& Samuel, H.\ 2013, Earth and Planetary Science Letters, 366, 6 

\bibitem[Morishima et al.(2008)]{2008ApJ...685.1247M} Morishima, R., 
Schmidt, M.~W., Stadel, J., \& Moore, B.\ 2008, \apj, 685, 1247 



\bibitem[Nimmo 
\& Kleine(2007)]{2007Icar..191..497N} Nimmo, F., \& Kleine, T.\ 2007, \icarus, 191, 497 


\bibitem[O'Brien et al.(2006)]{2006Icar..184...39O} O'Brien, D.~P., 
Morbidelli, A., \& Levison, H.~F.\ 2006, \icarus, 184, 39 

\bibitem[Papaloizou 
\& Nelson(2003)]{2003MNRAS.339..983P} Papaloizou, J.~C.~B., \& Nelson, R.~P.\ 2003, \mnras, 339, 983 


\bibitem[Petit et al.(2001)]{2001Icar..153..338P} Petit, J.-M., Morbidelli, 
A., \& Chambers, J.\ 2001, \icarus, 153, 338 

\bibitem[Pierens \& Raymond(2011)]{2011A&A...533A.131P} 
Pierens, A., \& Raymond, S.~N.\ 2011, \aap, 533, A131 



\bibitem[Raymond et al.(2009)]{2009Icar..203..644R} Raymond, S.~N., 
O'Brien, D.~P., Morbidelli, A., \& Kaib, N.~A.\ 2009, \icarus, 203, 644 
\bibitem[Raymond et al.(2004)]{2004Icar..168....1R} Raymond, S.~N., Quinn, 
T., \& Lunine, J.~I.\ 2004, \icarus, 168, 1
\bibitem[Raymond et al.(2005)]{2005ApJ...632..670R} Raymond, S.~N., Quinn, 
T., \& Lunine, J.~I.\ 2005, \apj, 632, 670 
\bibitem[Raymond et al.(2006)]{2006Icar..183..265R} Raymond, S.~N., Quinn, 
T., \& Lunine, J.~I.\ 2006, \icarus, 183, 265
\bibitem[Raymond et al.(2007)]{2007AsBio...7...66R} Raymond, S.~N., Quinn, 
T., \& Lunine, J.~I.\ 2007, Astrobiology, 7, 66 

\bibitem[Schlichting et al.(2012)]{2012ApJ...752....8S} Schlichting, H.~E., 
Warren, P.~H., \& Yin, Q.-Z.\ 2012, \apj, 752, 8 

\bibitem[Thommes et al.(2008)]{2008ApJ...676..728T} Thommes, E., Nagasawa, 
M., \& Lin, D.~N.~C.\ 2008, \apj, 676, 728 
\bibitem[Touboul et al.(2007)]{2007Natur.450.1206T} Touboul, M., Kleine, 
T., Bourdon, B., Palme, H., \& Wieler, R.\ 2007, \nat, 450, 1206 
\bibitem[Tsiganis et al.(2005)]{2005Natur.435..459T} Tsiganis, K., Gomes, 
R., Morbidelli, A., \& Levison, H.~F.\ 2005, \nat, 435, 459 

\bibitem[Usui et 
al.(2012)]{2012E&PSL.357..119U} Usui, T., Alexander, C.~M.~O., Wang, J., Simon, J.~I., \& Jones, J.~H.\ 2012, Earth and Planetary Science Letters, 357, 119 



\bibitem[Walsh et al.(2011)]{2011Natur.475..206W} Walsh, K.~J., Morbidelli, 
A., Raymond, S.~N., O'Brien, D.~P., \& Mandell, A.~M.\ 2011, Nature, 475, 206 

\bibitem[Wetherill(1996)]{1996Icar..119..219W} Wetherill, G.~W.\ 1996, 
\icarus, 119, 219 

\bibitem[Yin et al.(2002)]{2002Natur.418..949Y} Yin, Q., Jacobsen, S.~B., 
Yamashita, K., et al.\ 2002, \nat, 418, 949 







\end{thebibliography}
\end{document}